\documentclass[onecolumn]{IEEEtran}
\usepackage[latin1]{inputenc}
\usepackage[english]{babel}
\usepackage[margin=1in]{geometry}

\usepackage{blindtext}
\usepackage{amssymb}
\usepackage{hyperref}
\hypersetup{
    colorlinks=true,
    linkcolor=blue,
    filecolor=magenta,      
    urlcolor=cyan,
}
 
\usepackage{amsthm}
\usepackage[normalem]{ulem}
\usepackage{amsmath}
\usepackage{booktabs}
\usepackage{array}
\usepackage{cite}
\usepackage{soul}
\newtheorem{theorem}{Theorem}[section]
\newtheorem{corollary}{Corollary}[section]
\newtheorem{lemma}[theorem]{Lemma}
\newtheorem{proposition}[theorem]{Proposition}

\newtheorem{remark}{Remark}[section]
\newtheorem{conjecture}{Conjecture}

\newcommand\nc\newcommand
\nc{\cA}{\mathcal{A}}\nc{\cB}{\mathcal{B}}\nc{\cC}{\mathcal{C}}\nc{\cD}{\mathcal{D}}
\nc{\cE}{\mathcal{E}}\nc{\cF}{\mathcal{F}}\nc{\cG}{\mathcal{G}}\nc{\cH}{\mathcal{H}}
\nc{\cI}{\mathcal{I}}\nc{\cJ}{\mathcal{J}}\nc{\cK}{\mathcal{K}}\nc{\cL}{\mathcal{L}}
\nc{\cM}{\mathcal{M}}\nc{\cN}{\mathcal{N}}\nc{\cO}{\mathcal{O}}\nc{\cP}{\mathcal{P}}
\nc{\cQ}{\mathcal{Q}}\nc{\cR}{\mathcal{R}}\nc{\cS}{\mathcal{S}}\nc{\cT}{\mathcal{T}}
\nc{\cU}{\mathcal{U}}\nc{\cV}{\mathcal{V}}\nc{\cW}{\mathcal{W}}\nc{\cX}{\mathcal{X}}
\nc{\cY}{\mathcal{Y}}\nc{\cZ}{\mathcal{Z}}

\nc{\bba}{\mathbf{a}}\nc{\bbb}{\mathbf{b}}\nc{\bbc}{\mathbf{c}}\nc{\bbd}{\mathbf{d}}
\nc{\bbe}{\mathbf{e}}\nc{\bbf}{\mathbf{f}}\nc{\bbg}{\mathbf{g}}\nc{\bbh}{\mathbf{h}}
\nc{\bbi}{\mathbf{i}}\nc{\bbj}{\mathbf{j}}\nc{\bbk}{\mathbf{k}}\nc{\bbl}{\mathbf{l}}
\nc{\bbm}{\mathbf{m}}\nc{\bbn}{\mathbf{n}}\nc{\bbo}{\mathbf{o}}\nc{\bbp}{\mathbf{p}}
\nc{\bbq}{\mathbf{q}}\nc{\bbr}{\mathbf{r}}\nc{\bbs}{\mathbf{s}}\nc{\bbt}{\mathbf{t}}
\nc{\bbu}{\mathbf{u}}\nc{\bbv}{\mathbf{v}}\nc{\bbw}{\mathbf{w}}\nc{\bbx}{\mathbf{x}}
\nc{\bby}{\mathbf{y}}\nc{\bbz}{\mathbf{z}}

\nc{\bbA}{\mathbf{A}}\nc{\bbB}{\mathbf{B}}\nc{\bbC}{\mathbf{C}}\nc{\bbD}{\mathbf{D}}
\nc{\bbE}{\mathbf{E}}\nc{\bbF}{\mathbf{F}}\nc{\bbG}{\mathbf{G}}\nc{\bbH}{\mathbf{H}}
\nc{\bbI}{\mathbf{I}}\nc{\bbJ}{\mathbf{J}}\nc{\bbK}{\mathbf{K}}\nc{\bbL}{\mathbf{L}}
\nc{\bbM}{\mathbf{M}}\nc{\bbN}{\mathbf{N}}\nc{\bbO}{\mathbf{O}}\nc{\bbP}{\mathbf{P}}
\nc{\bbQ}{\mathbf{Q}}\nc{\bbR}{\mathbf{R}}\nc{\bbS}{\mathbf{S}}\nc{\bbT}{\mathbf{T}}
\nc{\bbU}{\mathbf{U}}\nc{\bbV}{\mathbf{V}}\nc{\bbW}{\mathbf{W}}\nc{\bfX}{\mathbf{X}}
\nc{\bbY}{\mathbf{Y}}\nc{\bbZ}{\mathbf{Z}}

\nc{\sA}{\mathsf{A}}\nc{\sB}{\mathsf{B}}\nc{\sC}{\mathsf{C}}\nc{\sD}{\mathsf{D}}
\nc{\sE}{\mathsf{E}}\nc{\sF}{\mathsf{F}}\nc{\sG}{\mathsf{G}}\nc{\sH}{\mathsf{H}}
\nc{\sI}{\mathsf{I}}\nc{\sJ}{\mathsf{J}}\nc{\sK}{\mathsf{K}}\nc{\sL}{\mathsf{L}}
\nc{\sM}{\mathsf{M}}\nc{\sN}{\mathsf{N}}\nc{\sO}{\mathsf{O}}\nc{\sP}{\mathsf{P}}
\nc{\sQ}{\mathsf{Q}}\nc{\sR}{\mathsf{R}}\nc{\sS}{\mathsf{S}}\nc{\sT}{\mathsf{T}}
\nc{\sU}{\mathsf{U}}\nc{\sV}{\mathsf{V}}\nc{\sW}{\mathsf{W}}\nc{\sX}{\mathsf{X}}
\nc{\sY}{\mathsf{Y}}\nc{\sZ}{\mathsf{Z}}

\newcommand{\mathset}[1]{\left\{#1\right\}}

\newcommand{\abs}[1]{\left|#1\right|}

\newcommand{\floorenv}[1]{\left\lfloor #1 \right\rfloor}
\newcommand{\parenv}[1]{\left( #1 \right)}
\newcommand{\sparenv}[1]{\left[ #1 \right]}

\nc{\set}[1]{\llbracket #1 \rrbracket}

\newcommand{\bal}[1]{\begin{align}\label{#1}}
\newcommand{\eal}{\end{align}}

\renewcommand{\leq}{\leqslant}

\renewcommand{\geq}{\geqslant}

\renewcommand{\Bbb}{\mathbb}

\renewcommand{\Bbb}{\mathbb}

\newcommand{\N}{{\Bbb N}}

\newcommand{\E}{{\Bbb E}}

\DeclareMathOperator{\supp}{Supp}

\newcommand{\ms}[1]{\{\{ #1 \}\}}

\theoremstyle{definition}
\newtheorem{definition}{Definition}[section]

\newcommand{\qbinom}[2]{\genfrac{[}{]}{0pt}{}{#1}{#2}}

\title{Bounds and Constructions for Generalized Batch Codes}

\date{\today} 

\author{Xiangliang Kong
\thanks{Xiangliang Kong is with the Department of Electrical Engineering-Systems, Tel Aviv University, Tel Aviv-Yafo 6997801, Israel ({e-mail: rongxlkong@gmail.com}). This work was supported by the European Research
Council (ERC) under Grant 852953.}
Ohad Elishco
\thanks{Ohad Elishco is with the School of Electrical and Computer Engineering, Ben-Gurion University of the Negev, Beer Sheva 8410501, Israel (e-mail: elishco@gmail.com).}
}

\begin{document}

\maketitle
\begin{abstract}
    Private information retrieval (PIR) codes and batch codes are two important types of codes that are designed for coded distributed storage systems and private information retrieval protocols. These codes have been the focus of much attention in recent years, as they enable efficient and secure storage and retrieval of data in distributed systems. 
    
    In this paper, we introduce a new class of codes called \emph{$(s,t)$-batch codes}. These codes are a type of storage codes that can handle any multi-set of $t$ requests, comprised of $s$ distinct information symbols. Importantly, PIR codes and batch codes are special cases of $(s,t)$-batch codes. 
    
    The main goal of this paper is to explore the relationship between the number of redundancy symbols and the $(s,t)$-batch code property. Specifically, we establish a lower bound on the number of redundancy symbols required and present several constructions of $(s,t)$-batch codes. Furthermore, we extend this property to the case where each request is a linear combination of information symbols, which we refer to as \emph{functional $(s,t)$-batch codes}. Specifically, we demonstrate that simplex codes are asymptotically optimal functional $(s,t)$-batch codes, in terms of the number of redundancy symbols required, under certain parameter regime. 
\end{abstract}

\section{Introduction}

Batch codes were originally developed by Ishai et al. \cite{IKOS04} for use in large-scale distributed storage systems and private information retrieval protocols. In this context, a batch code encodes $n$ information symbols, $x_1,\ldots, x_n$, into $N$ code symbols that are then distributed across $m$ separate storage buckets, which could represent servers, disks, or other virtual entities. A batch code is designed such that any multi-set of $t$ indices, ${i_1,\ldots,i_t}\subseteq [n]$, can be supported by reading at most one (and at most $a$ in general) code symbol from each bucket. As a result, a batch code can handle multiple requests for one or more (up to $t$) information symbols from different users simultaneously, a property referred to as \emph{availability}. This availability property is critical for achieving high throughput in distributed storage systems and has been extensively studied for other storage codes, such as locally repairable codes (LRCs), as seen in \cite{TB14,WZ14,HYUS15,RPDV16,TBF16,CMST20}.

Over the years, there have been many works exploring different variants of batch codes. One such variant, known as \emph{private information retrieval (PIR) codes}, was introduced in \cite{FVY15}. PIR codes are a special kind of batch code that requires every information symbol to have $t$ mutually disjoint recovering sets. This requirement can be satisfied by taking $i_1=i_2=\dots=i_t$ as the $t$ parallel requests under the batch code setting. Bounds and constructions for PIR codes have been explored in several papers, including \cite{FVY15,LR17,VRK17,AY19,KY21,HPPVY21}. Another type of batch code, known as \emph{combinatorial batch codes}, arises when the symbols stored in the buckets are simply copies of the information symbols. These codes have also been extensively studied, with notable works including \cite{PSW09,BRR12,SG16,ST20,CKZ21}. A special class of batch codes with $t=n$, called \emph{switch codes}, has been explored in \cite{WSCB13,WKC15,CGHZ15,BCSY18} to facilitate data routing in network switches. Recently, Zhang et al. \cite{ZEY20} (with related work in \cite{YY21}) investigated yet another variant of batch codes called \emph{functional batch codes}, where the $t$ parallel requests can be linear combinations of the information symbols.

Among all these works about different variants of batch codes, the central question is minimizing the number of servers $m$ while maximizing the rate of the code $\rho=n/N$ for a given number of information symbols $n$ and a required number of parallel requests $t$. As the base case, the batch codes where each server stores one symbol (i.e.,  $m=N$) tends to be particularly useful for constructing general batch codes. This class of batch codes are called \emph{primitive} batch codes and is the most studied one. For this case, the above central question is simplified as looking for the trade-off between the redundancy $N-n$ and the number of parallel requests $t$. In this paper, we focus on the primitive case and if there is no confusion, we use batch codes to refer to the primitive batch codes for simplicity. 

In their initial work, Ishai et al. \cite{IKOS04} used unbalanced expanders, subcube codes, smooth codes, and Reed-Muller codes to create batch codes for a constant rate $\rho<1$. Rawat et al. \cite{RSDG16} used graph theory to construct batch codes that achieve asymptotically optimal rates of $1-o_t(1)$. Vardy and Yaakobi \cite{VY16} then built batch codes with redundancy of $O(\sqrt{n}\log(n))$ for any fixed $t$, which is nearly optimal based on the redundancy lower bound of $O(\sqrt{n})$ obtained by \cite{RVW22} and \cite{LW21batch}. In \cite{AY19}, the authors 
obtained batch codes with smaller redundancy than those in \cite{RSDG16} through multiplicity codes. Using tools from finite geometry, \cite{PPV20} 
constructed batch codes with the smallest known redundancy for $t=O(n^{1/4})$. Recently, Li and Wooters \cite{LW21} introduced lifted multiplicity codes and showed that these codes are PIR codes with the best-known trade-off between redundancy and the number of requests $t=n^{\varepsilon}$ for any constant $0<\varepsilon<1/2$. In \cite{HPPVY21}, the authors considered lifted multiplicity codes for the multivariate case and obtained batch codes and PIR codes with the smallest known redundancy for some other parameter regimes. We refer the interested reader to Table II - Table V in \cite{HPPVY21}, which provide the trade-off between redundancy and the number of requests of both PIR codes and batch codes.

To further study the difference between PIR codes and batch codes, we introduce a new family of linear codes, dubbed \emph{generalized batch codes}. We first give a non-formal definition of $(s,t)$-batch codes for integers $1\leq s\leq t$. An $(s,t)$-batch code encodes $n$ information symbols $x_1,\ldots, x_n$ into $N$ code symbols, such that any $t$ parallel requests of at most $s$ information symbols $x_{i_1},\ldots,x_{i_s}$ can be supported by reading each code symbol at most once. This new class of codes is a hybrid between PIR codes and batch codes, and it caters to the intermediate type of requests. The batch code setting arises when $t$ different users hold queries $i_j$ and wish to retrieve data from devices. Usually in practice, there is a small amount of data, known as the ``hot'' data, that needs to be accessed much more frequently than the other data. Therefore, it is reasonable to consider a multi-set of requests consisting of only a few distinct queries. From this perspective, $(s,t)$-batch codes might be more practical than batch codes when $s$ is much smaller than $t$. 

Following similar lines as previous works, our focus in this paper is on the asymptotic behavior of redundancy with respect to the total number of information symbols $n$, the number of parallel requests $t$, and the number of distinct queries $s$. Our results consist of three parts. First, we prove a lower bound on the redundancy of $(s,t)$-batch codes, which slightly improves upon the previous lower bound for PIR codes in \cite{RVW22} and matches the result for batch codes in \cite{LW21batch}. 
Second, we provide two constructions of $(s,t)$-batch codes. The first is a recursive construction that shows that for constant $s$, the redundancy required by an $(s,t)$-batch code is only the redundancy required by $t$-PIR codes times $(O\log{n})$. The second is a modification of the random construction based on finite geometries in \cite{PPV20}, which demonstrates that for $t=o(n^{\frac{1}{3}})$ and $s=O(\frac{t}{\log{n}})$, there exist $(s,t)$-batch codes with redundancy $O(t^{\frac{3}{2}})\sqrt{n}$. Finally, we explore the $(s,t)$-batch code property for functional batch codes and provide asymptotically optimal constructions for small $s$ using simplex codes.

The rest of the paper is structured as follows. Section \ref{sec:def}  presents the formal definition of generalized $(s,t)$-batch codes and discusses some fundamental properties of these codes. In Section \ref{sec:bounds}, we prove a general lower bound on the redundancy of $(s,t)$-batch codes. Section 4 contains several constructions of $(s,t)$-batch codes, which provide upper bounds on the minimum redundancy. In Section 5, we explore the trade-off between redundancy and functional $(s,t)$-batch code property. Finally, in Section 6, we conclude the paper by highlighting some unresolved issues.

\section{Definitions and Notations}
\label{sec:def}

For positive integers $n>m$, we denote by $[m,n]$ the set of integers $\mathset{m,m+1,\dots,n}$ and we denote by $[n]$ the set $\mathset{1,2,\dots,n}$. For an integer $s>0$, we denote by ${[n]\choose s}$ the collection of all subsets of $[n]$ of size $s$. Let $i_1,\dots,i_s\in [n]$, not necessarily distinct, and let $a_1,\dots,a_s\in \N$. We define   
\begin{align}\label{eq_legalrequest}
I:=\ms{\underbrace{i_1,\ldots,i_1}_{a_1},\underbrace{i_2,\ldots,i_2}_{a_2},\ldots, \underbrace{i_s,\ldots,i_s}_{a_s}} 
\end{align} 
as the multi-set consisting of $i_1,\ldots,i_s$ with multiplicity $a_1,\ldots,a_s$, respectively. 
For simplicity, we will also use the notation $I=\ms{i_1^{(a_1)},i_2^{(a_2)},\ldots, i_s^{(a_s)}}$. 
For a multi-set $I=\ms{i_1^{(a_1)},i_2^{(a_2)},\ldots, i_s^{(a_s)}}$, we denote $I\subseteq [n]$ if $i_j\in [n]$ for every $1\leq j\leq s$. 
Let $\mathcal{C}$ be a linear code over field $\mathbb{F}$, and assume it encodes $n$ information symbols $x_1,\ldots,x_n$ into $N$ coded symbols $y_1,\ldots,y_N$. Then $\mathcal{C}$ is called \emph{systematic} if $y_i=x_i$ for $i\in [n]$. 
For a codeword $\bbc\in \mathcal{C}$ and a set of indices $R\subseteq [N]$, let $\mathbf{c}|_{R}$ be the vector obtained by projecting the coordinates of $\bbc$ to $R$, and defined $\mathcal{C}|_{R}=\{\mathbf{c}|_{R}: \mathbf{c}\in \mathcal{C}\}$. 
Moreover, for a vector $\mathbf{v}\in \mathbb{F}^N$, we denote $\supp(\mathbf{v})$ as the support set of $\mathbf{v}$. 
We now introduce the definition of $(s,t)$-batch codes.

\begin{definition}\label{GeneralBatch}
For a field $\mathbb{F}$, let $\mathcal{C}\subseteq \mathbb{F}^{N}$ be a systematic linear code of dimension $n$. The code $\mathcal{C}$ is an \textbf{$(s,t)$-batch code} if for any multi-set of indices $I=\ms{i_1^{(a_1)},i_2^{(a_2)},\ldots, i_{s}^{(a_{s})}}\subseteq [n]$ with $\sum_{j=1}^{s}a_j=t$, where $i_1,\dots,i_s\in [n]$ are not necessarily distinct, there are $t$ pairwise disjoint sets 
$$R_{1,1},\ldots,R_{1,a_1},\ldots, R_{s,1},\ldots,R_{s,a_{s}}\subseteq [N]$$
such that for all $\mathbf{c}\in \mathcal{C}$, $j\in [s]$ and $l\in [a_j]$, $\mathbf{c}({i_j})$ is an $\mathbb{F}$-linear combination of codeword symbols in $\mathbf{c}|_{R_{j,l}}$. Moreover, we call $R_{j,l}$ a \textbf{recovering} set of $i_j$. 
\end{definition}
Generally speaking, an $(s,t)$-batch code is a code for which any set of $s$ (information) symbols can be recovered simultaneously from multiple sets of other code symbols. 
Notice that in the definition above, the multi-set $I \subseteq [n]$ only involves information symbols. 


\begin{remark}
 By Definition \ref{GeneralBatch}, a $t$-PIR code is a $(1,t)$-batch code, and a $t$-batch code is a $(t,t)$-batch code. Clearly, a $t$-batch code is always an $(s,t)$-batch code (since $s\leq t$).   
\end{remark}

We denote by $r(n;s,t)$ the minimum redundancy of an $(s,t)$-batch code that encodes $n$ information symbols, then for every 
$1\leq s\leq t$ we immediately obtain the inequality  
\begin{equation}\label{eq1}
   r(n;1,t)\leq r(n;s,t)\leq r(n;t,t).
\end{equation}
In all of the redundancy notation, if $n$ is clear from the context, we omit $n$ from the notation and write $r(s,t)$ 
instead of $r(n;s,t)$.

Throughout the paper, we use Bachmann-Landau notations to denote the asymptotic upper and lower bounds of a function. Formally, let $f(n)$ and $g(n)$ be two non-negative functions defined on the positive integers. We say that $f=O(g)$ if there exist positive constants $c$ and $n_0$ such that $f(n)\leq cg(n)$ for all $n\geq n_0$ and we say that $f=o(g)$ if for every $\epsilon>0$ there exists a constant $n_0$ such that $f(n)\leq \epsilon g(n)$ for all $n\geq n_0$. 


\section{Lower bounds on redundancy of $(s,t)$-batch codes} 
\label{sec:bounds}
In this section, we provide a general lower bound on $r(s,t)$ using similar techniques as in \cite{AG21}, modified to suit our problem. 
%
%
%
%
To bound below the redundancy $r(s,t)$ we first introduce a slightly different family of codes, dubbed $(u,v)$-ordered batch codes, which is a generalization of the ordered batch codes introduced in \cite{AG21}. The main difference between ordered batch codes and "regular" batch codes is an additional structure that allows ordered recovery of symbols. The reason for defining ordered batch codes is that their redundancy can be estimated more easily. We then describe a relation between batch codes and ordered batch codes and leverage this relation to bound below the redundancy of "regular" batch codes. 
Thus, we start with the definition of $(u,v)$-ordered batch codes for $u,v\in \N$. 

\begin{definition}\label{G_orderedbatch}
For a field $\mathbb{F}$, let $\mathcal{C}\subseteq \mathbb{F}^{N}$ be a systematic linear code of dimension $n$. For $u,v\in \mathbb{N}$, $\mathcal{C}$ is called a \textbf{$(u,v)$-ordered-batch code} if for any set of $u$ indices $I=\{i_1,\ldots,i_u\}\subseteq[n]$, there are $uv$ pairwise disjoint sets 
$$R_{(1,1)},\ldots,R_{(1,v)},\ldots, R_{(u,1)},\ldots,R_{(u,v)}\subseteq [N]$$
such that for all $\mathbf{c}\in \mathcal{C}$, $j\in [u]$ and $l\in [v]$, $\mathbf{c}({i_j})$ is an $\mathbb{F}$-linear combination of codeword symbols in $\mathbf{c}|_{R_{j,l}}$. Moreover, $R_{j,l}$'s satisfy the following additional property: Define the directed graph $D_I$ with vertex set $I$ and edges $i_j\rightarrow i_k$ if $i_k\in \bigcup_{l\in [v]}R_{j,l}$, then the graph $D_I$ is a directed acyclic graph (DAG).
\end{definition}

An intuitive explanation for the definition is as follows: For every set of $u$ indices, there are $uv$ mutually disjoint recovering sets, such that every index has $v$ recovering sets. The additional property implies an order of recovery, i.e., if $D_I$ is a DAG, it can be topologically ordered (see, for example, \cite{ban2001digraphs}). This means, that the additional requirement guarantees that there is an order of recovery, with which it will be possible to recover all the required symbols no matter the recovering sets we choose. 

Next, we connect $(s,t)$-batch codes to $(u,v)$-ordered batch codes.
\begin{proposition}\label{prop1}
If a linear code $\mathcal{C}\subseteq \mathbb{F}^{N}$ is an $(s,t)$-batch code, then it is a $(u,v)$-ordered-batch code for every $1\leq u\leq s$ and $1 \leq v\leq \lfloor\frac{t}{u}\rfloor-1$. 
\end{proposition}

\begin{IEEEproof}
Let $\mathcal{C}$ be an $(s,t)$-batch code. By  definition, a $(u,v)$-ordered-batch code is a  $(u',v')$-ordered-batch code for $u'\leq u$ and $v'\leq v$. Therefore, we only need to show that $\mathcal{C}$ is a $(u,\lfloor\frac{t}{u}\rfloor-1)$-ordered-batch code for every $1\leq u\leq s$. We prove only the case when $u=s$ but the rest of the cases are proved similarly. 

Let $I=\{i_1,\ldots,i_s\}\subseteq [n]$ be a set of $s$ distinct indices. By Definition \ref{GeneralBatch}, for every position $i_j$, $j\in [s]$, there are $s\cdot \lfloor\frac{t}{s}\rfloor$ mutually disjoint recovering sets 
\begin{equation*}
    R_{(1,1)},\ldots,R_{(1,\lfloor\frac{t}{s}\rfloor)},\ldots,R_{(s,1)},\ldots,R_{(s,\lfloor\frac{t}{s}\rfloor)},
\end{equation*}
such that for every element $i_j\in I$, there are $\lfloor\frac{t}{s}\rfloor$ disjoint recovering sets $R_{(j,1)},\ldots,R_{(j,\lfloor\frac{t}{s}\rfloor)}$ satisfying for all $\mathbf{c}\in \mathcal{C}$ and $1\leq l\leq \lfloor\frac{t}{s}\rfloor$, $\mathbf{c}({i_j})$ is an $\mathbb{F}$-linear combination of codeword symbols in $\mathbf{c}|_{R_{(j,l)}}$.

Now, consider the directed graph $D_I$ with $I$ as vertex set and $i_j\rightarrow i_k$ if $i_k\in \bigcup_{l=1}^{\lfloor\frac{t}{s}\rfloor}R_{j,l}$. In order to obtain an ordered batch code, we might need to alter this graph (and accordingly, the recovering sets). Since the recovering sets are mutually disjoint, for each $i_j\in I$, its in-degree is at most $1$. 
This indicates that the directed cycles in $D_I$ are vertex disjoint. To see this, assume towards a contradiction that there are two directed cycles $C_1,C_2$ that share a common vertex $i$. Since the in-degree of every vertex is at most one, the predecessor of $i$ in both cycles must be the same. By repeating this argument, we obtain that $C_1=C_2$, contradicting our assumption. 
Since all the directed cycles in $D_I$ are vertex disjoint, we can remove a collection of vertex-disjoint edges to make $D_I$ a DAG. Note that for every $i_j\in I$ and a directed edge starting from $i_j$ in $D_I$, there is a unique recovering set for $i_j$ that corresponds to this edge. Thus, we can remove at most one recovering set from each $i_j\in I$ to make the remaining graph a DAG.
\end{IEEEproof}

The following theorem provides a lower bound on the redundancy of $(u,v)$-ordered-batch codes. 

\begin{theorem}\label{thm1}
Let $\mathcal{C}\subseteq\mathbb{F}_q^{N}$ be a $(u,v)$-ordered-batch code ($u,v\in \mathbb{N}$) of dimension $n$ and redundancy $r$ over a field of size $q$. Assume that $q\geq \max\{4u{n\choose u},7\}$, $v\geq 2$, and that $uv\leq n$. Then 
\begin{equation}\label{eq3}
    {{r+2u+1-v}\choose 2u}+\sum_{i=1}^{u}{{r+2u+1-(v+i)}\choose {2u-i}}{{v+i-3}\choose i}\geq {n\choose u}.
\end{equation}
\end{theorem} 

Before proving the theorem, we give an overview of the proof's idea, which generalizes the idea in \cite{RVW22} for the case when $u=1$ and $v=2$. 
We briefly explain the core idea that appears in \cite{RVW22} since the generalization relies on a similar concept. 
Consider $\mathcal{C}$, a $(1,2)$-ordered-batch code. Then every $i\in[n]$ has $2$ mutually disjoint recovering sets $R_{i,1},R_{i,2}\subseteq[N]\setminus \{i\}$. In other words, for each $i\in[n]$, there are 2 codewords $\mathbf{c}_{i,1},\mathbf{c}_{i,2}\in\mathcal{C}^{\perp}$ such that
$$i\in \supp(\mathbf{c}_{i,j})\subseteq R_{i,j}\cup\{i\},~j=1,2.$$
Denote $r$ as the redundancy of $\mathcal{C}$ and assume that $G\in \mathbb{F}_q^{r\times N}$ is the generator matrix of $\mathcal{C}^{\perp}$ with $\mathbf{w}_i$ as the $i$-th column. Denote $\mathbf{c}_{i,1}=\alpha_i\cdot G$ and $\mathbf{c}_{i,2}=\beta_i\cdot G$. Then, one can construct a collection of $r$-variate polynomials $\{P_i(X)\}_{i\in [n]}$ by defining 
\begin{equation}\label{idea_eq1}
    P_i(X)=\langle\alpha_i,X\rangle\langle\beta_i,X\rangle,
\end{equation}
where $X=(x_1,\ldots,x_r)$. By taking $X=\mathbf{w}_i$, one can verify that $\{P_i(X)\}_{i\in [n]}$ are linearly independent. On the other hand, each $P_i$ is an $r$-variate polynomial of degree $2$. Thus, we have $n\leq {r+1\choose 2}$, which leads to the lower bound on $r$. In \cite{AG21}, Alrabiah and Guruswami extended the above proof for the case when $\mathcal{C}$ is a $(u,2)$-ordered-batch code. Using the DAG property of the ordered batch code, they managed to construct a collection of linearly independent $r$-variate polynomial $\{P_I(X)\}_{I\in {[n]\choose u}}$. 

Note that for $v\geq 3$, $(u,v)$-ordered-batch codes  naturally requires more redundancy than $(u,2)$-ordered-batch codes. Therefore, one expects a larger lower bound for $(u,v)$-ordered-batch code than that for $(u,2)$-ordered-batch code. When each $i\in [n]$ has many mutually disjoint recovering sets, we manage to show that there are $\mathbf{c}_{i,1},\mathbf{c}_{i,2}\in\mathcal{C}^{\perp}$ satisfying $\supp(\mathbf{c}_{i,1})\cap \supp(\mathbf{c}_{i,2})=\{i\}$ and $\mathbf{c}_{i,2}$ lies in a low-dimensional subspace $V$ of $\mathcal{C}^{\perp}$ (see Lemma \ref{lem1.1}). Then, using similar arguments as \cite{AG21}, we can construct a collection of linearly independent $r$-variate polynomial $\{P_I(X)\}_{I\in {[n]\choose u}}$. With the extra property that $\mathbf{c}_{i,2}$ lies in a low-dimensional subspace of $\mathcal{C}^{\perp}$, we can obtain extra conditions on the degree of the collection of polynomials $\{P_I(X)\}_{I\in {[n]\choose u}}$. This enables us to get a slightly better bound for $(u,v)$-ordered-batch code than that proved in \cite{AG21}.

Next, we delve into the details of the proof. First, we prove the following lemma which guarantees the existence of the low-dimensional subspace of $\mathcal{C}^{\perp}$ described above.

\begin{lemma}\label{lem1.1}
Let $\mathcal{C}$ be the linear $(u,v)$-ordered-batch code defined in Theorem \ref{thm1} and denote by $\mathcal{C}^{\perp}$ the dual code of $\mathcal{C}$. 
Then, there exists an $(r-v+2)$-dimensional subspace $V$ of $\mathcal{C}^{\perp}$ such that for every  $I=\{i_1,\ldots,i_u\}\subseteq[n]$ and every $i_j\in I$, $V$ contains a vector $\mathbf{v}_j$ with $i_j\in \supp(\mathbf{v}_j)\subseteq (\bigcup_{l=2}^{v}R_{j,l})\cup \{i_j\}$, where $R_{j,1},\ldots, R_{j,v}$ are the $v$ mutually disjoint recovering sets of $i_j$ (notice that the union starts from $l=2$.). 
\end{lemma}

\begin{IEEEproof}
By the definition of $(u,v)$-ordered-batch code, for every $I=\{i_1,\ldots,i_u\}\subseteq [n]$ and every $i_j\in I$, there are $v$ mutually disjoint recovering sets $\{R_{j,l}\}_{1\leq l\leq v}$ of $i_j$ in $[N]$. In other words, we can find $v$ codewords $\{\mathbf{c}_{j,l}\}_{1\leq l\leq v}$ in $\mathcal{C}^{\perp}$ such that $i_j\in \supp(\mathbf{c}_{j,l})\subseteq R_{j,l}\cup\{i_j\}$. Clearly, $\mathbf{c}_{j,1},\ldots,\mathbf{c}_{j,v}$ are linearly independent and $\text{Span}_{\mathbb{F}_q}\{\mathbf{c}_{j,l}\}_{l=2}^{v}$ is a $(v-1)$-dim subspace of $\mathcal{C}^{\perp}$. Based on this observation, we prove the existence of $V$ through a probabilistic argument. 

Pick an $(r-v+2)$-dim subspace $V$ from $\mathcal{C}^{\perp}$ randomly according to the uniform distribution. For every $(v-1)$-dim subspace $U$ in $\mathcal{C}^{\perp}$, we have $\dim(V\cap U)\geq 1$. Thus, for every $I=\{i_1,\ldots,i_u\}$ and every $i_j\in I$, $\dim(V\cap \text{Span}_{\mathbb{F}_q}\{\mathbf{c}_{j,l}\}_{l=2}^{v})\geq 1$. 
This implies that $V$ always contains a vector supported on $(\bigcup_{l=2}^{v}R_{j,l})\cup \{i_j\}$. To finish the proof, we show that with positive probability, $V\cap \text{Span}_{\mathbb{F}_q}\{\mathbf{c}_{j,l}\}_{l=2}^{v}$ contains a vector that is non-zero at coordinate $i_j$.

Fix a set $I=\{i_1,\ldots,i_u\}\subseteq [n]$. For every $j\in [u]$, denote $U_j(I)=\{\mathbf{c}\in \text{Span}_{\mathbb{F}_q}\{\mathbf{c}_{j,l}\}_{l=2}^{v}: \mathbf{c}(i_j)=0\}$. Then, $U_j(I)$ is a $(v-2)$-dim subspace in $\text{Span}_{\mathbb{F}_q}\{\mathbf{c}_{j,l}\}_{l=2}$. Define $A_j(I)$ as the event that 
$V\cap \text{Span}_{\mathbb{F}_q}\{\mathbf{c}_{j,l}\}_{l=2}^{v}\subseteq U_j(I)$. 
If we show that $\Pr(A_j(I))<1$, we obtain that with positive probability, $V$ contains a vector that is non-zero in the $i_j$ coordinate. 
Since $V\cap U_j(I)\subseteq V\cap \text{Span}_{\mathbb{F}_q}\{\mathbf{c}_{j,l}\}_{l=2}^{v}$, the event $A_j(I)$ occurs if and only if 
$$\dim(V\cap \text{Span}_{\mathbb{F}_q}\{\mathbf{c}_{j,l}\}_{l=2}^{v})=\dim(V\cap U_j(I))>0.$$
Since $\dim(U_j(I))=0$ when $v=2$, we have $\Pr(A_j(I))=0$ in this case. For $v\geq 3$, we have
\begin{align}
    \Pr(A_j(I))&=\sum_{w=1}^{v-2}\Pr(\dim(V\cap \text{Span}_{\mathbb{F}_q}\{\mathbf{c}_{j,l}\}_{l=2}^{v})=\dim(V\cap U_j(I))=w) \nonumber
\end{align}
For each $1\leq w\leq v-2$, by Lemma \ref{lem71}, we know that the number of $(r-v+2)$-dim subspace in $\mathcal{C}^{\perp}$ intersecting $\text{Span}_{\mathbb{F}_q}\{\mathbf{c}_{j,l}\}_{l=2}^{v}$ at a $w$-dim subspace in $U_j(I)$ is given by $q^{(v-1-w)(r-v+2-w)}\qbinom{r-(v-1)}{r-v+2-w}_q\qbinom{v-2}{w}_q$, where for positive integers $n\geq k\geq 1$, $\qbinom{n}{k}_q=\frac{(q^n-1)\dots(q^{n-k+1}-1)}{(q^k-1)\dots (q-1)}$. Since $V$ is chosen uniformly at random from $\mathcal{C}^{\perp}$, we have
\begin{align}
    \Pr(A_j(I))&= \sum_{w=1}^{v-2}\frac{q^{(v-1-w)(r-v+2-w)}\qbinom{r-(v-1)}{r-v+2-w}_q\qbinom{v-2}{w}_q}{\qbinom{r}{r-v+2}_q} \nonumber\\
    \label{eq4b} &\leq \sum_{w=1}^{v-2}\frac{1+\frac{4w}{q-1}+\frac{4w^2}{(q-1)^2}}{q^{w^2}},
\end{align}
where (\ref{eq4b}) follows from Lemma \ref{lem73}. By $q\geq 4u{n\choose u}\geq 4(v-2)$ and $q\geq 7$, (\ref{eq4b}) can be upper bounded by 
\begin{align*}
    \sum_{w=1}^{v-2}\frac{1+\frac{4w}{q-1}+\frac{4w^2}{(q-1)^2}}{q^{w^2}}&\leq \frac{1+\frac{4}{q-1}+\frac{4}{(q-1)^2}}{q}+\frac{v-2}{q^4}(1+\frac{4(v-2)}{q-1}+\frac{4(v-2)^2}{(q-1)^2})\\
    &\leq \frac{1+\frac{4}{q-1}+\frac{4}{(q-1)^2}}{q}+\frac{4}{q^3}< \frac{2}{q}.
\end{align*}
Thus, by the union bound, the probability that there exists a subset $I=\{i_1,\ldots,i_u\}\in{[n]\choose u}$ such that the event $A_j(I)$ occurs for some $j\in [u]$, is at most 
$$u\cdot {n\choose u}\cdot\Pr(A_j(I))\leq \frac{2u{n\choose u}}{q}<1.$$ 
Therefore, with positive probability, there is an $(r-v+2)$-dim subspace $V$ of $\mathcal{C}^{\perp}$ such that $V\cap \text{Span}_{\mathbb{F}_q}\{\mathbf{c}_{j,l}\}_{l=2}^{v}\nsubseteq U_j(I)$ for every $I=\{i_1,\ldots,i_u\}$ and every $i_j\in I$. Then, we can take $\mathbf{v}_j$ as any vector in $\text{Span}_{\mathbb{F}_q}\{\mathbf{c}_{j,l}\}_{l=2}^{v}\setminus U_j(I)$. Clearly, $i_j\in \supp(\mathbf{v}_j)\subseteq (\bigcup_{l=2}^{v}R_{j,l})\cup \{i_j\}$, which concludes the proof.
\end{IEEEproof}

Now, with the help of Lemma \ref{lem1.1}, we can move on to the proof of Theorem \ref{thm1}.

\begin{IEEEproof}[Proof of Theorem \ref{thm1}] 
Let $G\in \mathbb{F}_q^{r\times N}$ be the generator matrix of $\mathcal{C}^{\perp}$ and denote by $\mathbf{w}_{i}$ the $i$-th column of $G$. Then, for each $\mathbf{c}\in \mathcal{C}^{\perp}$, we have $$\mathbf{c}=\mathbf{\alpha}\cdot G=(\langle\mathbf{\alpha},\mathbf{w}_1\rangle,\ldots,\langle\mathbf{\alpha},\mathbf{w}_N\rangle)$$ for some vector $\alpha\in \mathbb{F}_q^{r}$.
From Lemma \ref{lem1.1} we obtain $V$, which is an $(r-v+2)$-dim subspace of $\mathcal{C}^{\perp}$, together with two vectors $\mathbf{c}_{j,1}$ and $\mathbf{v}_j\in V$. 
W.l.o.g., we can assume that $V$ is spanned by the first $r-v+2$ rows of $G$. 

Given $I=\{i_1,\ldots,i_u\}$, for every $j\in [u]$, let $\alpha_{j}$ and $\beta_{j}$ denote the vectors in $\mathbb{F}_q^{r}$ such that $\mathbf{c}_{j,1}=\alpha_{j}\cdot G$ and $\mathbf{v}_j=\beta_{j}\cdot G$.
Since $V$ spanned by the first $r-v+2$ first rows in $G$, we have $\beta_j(l)=0$ for all $r-v+3\leq l\leq r$. For $X=(x_1,\ldots,x_r)$ with $x_i$ being an indeterminate over $\mathbb{F}_q$, define the polynomial
\begin{equation}\label{eq5}
    P_{I}(X):=\prod_{j=1}^{u}\langle\alpha_j,X\rangle\langle\beta_j,X\rangle.
\end{equation}
By the definition of $\alpha_j$ and $\beta_j$, $\langle\alpha_j,X\rangle$ is an $\mathbb{F}_q$-linear combination of $x_1,\ldots,x_r$ and $\langle\beta_j,X\rangle$ is an $\mathbb{F}_q$-linear combination of $x_1,\ldots,x_{r-v+2}$. 
Note that each $\langle\alpha_j,X\rangle\langle\beta_j,X\rangle$ is a quadratic polynomial in $\mathbb{F}_q[x_1,\ldots,x_r]$ and $P_I$ is a product of $u$ such quadratic polynomials. 
Therefore, each $P_I$ is an $\mathbb{F}_q$-linear combination of monomials $x_1^{d_1}x_2^{d_2}\cdots x_{r}^{d_r}$ such that 
\begin{equation}\label{eq6}
\begin{cases}
   \sum_{i=1}^{r}d_i=2u,\\
   \sum_{i=1}^{r-v+2}d_i\geq u,\\
   d_i\geq 0 \text{~for~} 1\leq i\leq r.
\end{cases}
\end{equation}

For a fixed $0\leq i\leq u$, there are ${{2u-i+(r-v+1)}\choose {2u-i}}$ different choices of non-negative integers $d_1,\dots,d_{r-v+2}$ such that $\sum_{j=1}^{r-v+2} d_j=2u-i$. When $v\geq 3$, there are ${{i+(v-3)}\choose i}$ different choices of non-negative integers $d_{r-v+3},\dots,d_r$ for which $\sum_{j=r-v+3}^{r}d_j=i$ (see \cite[Proposition 1.5]{Jukna2011}). Notice that when $v=2$, (\ref{eq6}) reduces to
\begin{equation*}
\begin{cases}
   \sum_{i=1}^{r}d_i=2u,\\
   d_i\geq 0 \text{~for~} 1\leq i\leq r.
\end{cases}
\end{equation*}
Thus, when $v\geq 3$, there are $\sum_{i=0}^{u}{{2u-i+(r-v+1)}\choose {2u-i}}{{i+v-3}\choose i}$ different monomials $x_1^{d_1}x_2^{d_2}\cdots x_{r}^{d_r}$ satisfying (\ref{eq6}) and when $v=2$, there are ${{r+2u+1-v}\choose 2u}$ different such monomials. Setting ${i\choose j}=0$, if $i<j$, we can unify these two cases together and obtain that the number of monomials satisfying (\ref{eq6}) is 
\begin{equation}\label{eq6.5}
    {{r+2u+1-v}\choose 2u}+\sum_{i=1}^{u}{{r+2u+1-(v+i)}\choose {2u-i}}{{v+i-3}\choose i}.
\end{equation}
 
Next, we show that the collection of polynomials $\{P_{I}: I\in {[n]\choose u}\}$ are linearly independent. 
Since each $P_I$ is a linear combination of monomials $x_1^{d_1}x_2^{d_2}\cdots x_{r}^{d_r}$ satisfying (\ref{eq6}) and there are (\ref{eq6.5}) such possible monomials, $(\ref{eq3})$ follows from the linear independence of $\{P_{I}: I\in {[n]\choose u}\}$. 

Let $z_1,\ldots,z_N$ be $N$ variables over $\mathbb{F}_q$ and plug $X=\sum_{k=1}^{N}z_k\mathbf{w}_k$ into $P_{I}$. This gives us the following homogeneous polynomial
\begin{align}\label{eq7}
    Q_{I}(z_1,\ldots,z_N):=P_{I}(\sum_{k=1}^{N}z_k\mathbf{w}_k)&=\prod_{j=1}^{u}\langle\alpha_j,\sum_{k=1}^{N}z_k\mathbf{w}_k\rangle\langle\beta_j,\sum_{k=1}^{N}z_k\mathbf{w}_k\rangle \nonumber\\
    &=\prod_{j=1}^{u}\left(\sum_{k=1}^{N}z_k\langle\alpha_j,\mathbf{w}_k\rangle\right)\left(\sum_{k=1}^{N}z_k\langle\beta_j,\mathbf{w}_k\rangle\right).
\end{align}
Note that the polynomials $\{P_{I}: I\in {[n]\choose u}\}$ being linearly dependent will naturally lead to $\{Q_{I}: I\in {[n]\choose u}\}$ also being linearly dependent. Therefore, it suffices to show that polynomials in $\{Q_{I}: I\in {[n]\choose u}\}$ are linearly independent. For this purpose, we claim that for any $J\in {[n]\choose u}$, the monomial $\prod_{k\in J}z_k^2$ has a nonzero coefficient in $Q_I$ if and only if $J=I$. Then, linear independence follows naturally.

Assume first that the monomial $\prod_{k\in J}z_k^2$ has a nonzero coefficient in $Q_I$. Using (\ref{eq7}), since $\langle\beta_j,\mathbf{w}_k\rangle=\mathbf{v}_j(k)$, a nonzero coefficient $\langle\beta_j,\mathbf{w}_k\rangle$ implies that $k\in (\bigcup_{l=2}^{v}R_{j,l})\cup\{i_j\}$. Similarly, a nonzero coefficient $\langle\alpha_j,\mathbf{w}_k\rangle$ implies that $k\in R_{j,1}\cup\{i_j\}$. Since $\{R_{j,l}\}_{1\leq j\leq u \atop 1\leq l\leq v}$ are mutually disjoint, if $k\notin I$, the degree of $z_k$ in every term of $Q_{I}$ is at most $1$. Therefore, if the coefficient of the term $\prod_{k\in J}z_k^2$ is nonzero, we must have $k\in I$ for all $k\in J$, which leads to $J=I$. This proves one direction of the claim.

For the other direction, we assume $J=I$ and we show that the coefficient of the term $\prod_{k\in I}z_k^2$ is nonzero in $Q_{I}$. Recall that $Q_{I}(z_1,\ldots,z_N)= \prod_{j=1}^{u}\left(\sum_{k=1}^{N}z_k\langle\alpha_j,\mathbf{w}_k\rangle\right)\left(\sum_{k=1}^{N}z_k\langle\beta_j,\mathbf{w}_k\rangle\right)$ and note that for every $j\in [u]$ 
\begin{align*}
    \left(\sum_{k=1}^{N}z_k\langle\alpha_j,\mathbf{w}_k\rangle\right)\left(\sum_{k=1}^{N}z_k\langle\beta_j,\mathbf{w}_k\rangle\right)=\sum_{k,\tilde{k}\in [N]}\langle\alpha_j,\mathbf{w}_{k}\rangle\langle\beta_j,\mathbf{w}_{\tilde{k}}\rangle z_{k}z_{\tilde{k}}.
\end{align*}
Writing (\ref{eq7}) differently, we have
\begin{align*}
    Q_{I}(z_1,\ldots,z_N)
    &=\sum_{(k_1,\ldots,k_u)\in [N]^{u} \atop (\tilde{k}_1,\ldots,\tilde{k}_u)\in [N]^{u}}\left(\prod_{j=1}^{u}\langle\alpha_j,\mathbf{w}_{k_j}\rangle\langle\beta_j,\mathbf{w}_{\tilde{k}_j}\rangle\right)\prod_{j=1}^{u}z_{k_j}z_{\tilde{k}_j}.
\end{align*}
The coefficient of $\prod_{k\in I}z_k^2$ equals to the summation of  $\prod_{j=1}^{u}\langle\alpha_j,\mathbf{w}_{k_j}\rangle\langle\beta_j,\mathbf{w}_{\tilde{k}_j}\rangle$ over all $(k_1,\ldots,k_u)$, $(\tilde{k}_1,\ldots,\tilde{k}_u)\in [N]^{u}$ such that the multi-set  
\begin{equation}\label{eq7b}
\ms{k_1,\ldots,k_u,\tilde{k}_1,\ldots,\tilde{k}_u}=\ms{i_1^{(2)},i_2^{(2)},\ldots,i_u^{(2)}}.
\end{equation}
Next, we show that if $(k_1,\ldots,k_u)$, $(\tilde{k}_1,\ldots,\tilde{k}_u)\in [N]^{u}$ satisfy $\prod_{j=1}^{u}\langle\alpha_j,\mathbf{w}_{k_j}\rangle\langle\beta_j,\mathbf{w}_{\tilde{k}_j}\rangle\neq 0$ and (\ref{eq7b}), then
\begin{equation*}
    (k_1,\ldots,k_u)=(\tilde{k}_1,\ldots,\tilde{k}_u)=(i_1,\ldots,i_u),
\end{equation*}
implying that $J=I$. 


From $\prod_{j=1}^{u}\langle\alpha_j,\mathbf{w}_{k_j}\rangle\langle\beta_j,\mathbf{w}_{\tilde{k}_j}\rangle\neq 0$, we have $\langle\alpha_j,\mathbf{w}_{k_j}\rangle\langle\beta_j,\mathbf{w}_{\tilde{k}_j}\rangle\neq 0$ for every $1\leq j\leq u$.
Thus, $k_j\in R_{j,1}\cup\{i_j\}$ and $\tilde{k}_j\in \left(\bigcup_{l=2}^{v}R_{j,l}\right)\cup\{i_j\}$. 
Let $D_0$ be a directed graph with $I$ as a vertex set and directed edges $i_j\rightarrow k_j$ and $i_j\rightarrow \tilde{k}_j$. Specifically, when $k_j=i_j$ (or $\tilde{k}_j=i_j$), $i_j\rightarrow k_j$ (or $i_j\rightarrow \tilde{k}_j$) is considered as a self-loop in $D_0$ and when $k_j=\tilde{k}_j$, $i_j\rightarrow k_j$ and $i_j\rightarrow \tilde{k}_j$ are considered as two parallel directed edges in $D_0$. Since $R_{j,1},\ldots, R_{j,v}$ are mutually disjoint, $k_j=\tilde{k}_j$ iff $k_j=\tilde{k}_j=i_j$.
Because each $i_j$ appears twice in $\ms{k_1,\ldots,k_u,\tilde{k}_1,\ldots,\tilde{k}_u}$, the in-degree and the out-degree of every vertex in $D_0$ equals to $2$. By removing self-loops in $D_0$, we obtain a new directed graph $D_0'$ in which the in-degree and the out-degree of every vertex are equal. Therefore, $D_0'$ can be decomposed into a disjoint union of cycles. On the other hand, in the directed graph $D_I$ defined in Definition \ref{G_orderedbatch}, for $i_j,i_{j'}\in I$, there is a directed edge $i_j\rightarrow i_{j'}$ only if $i_{j'}\in \bigcup_{l=1}^{v}R_{j,l}$. Thus, $D_0'$ is a subgraph of $D_I$, which is a DAG. 
This implies that $D_0'$ is an empty graph. Therefore, $D_0$ contains only self-loops, which implies that $(k_1,\dots,k_u)=(\tilde{k}_1,\ldots,\tilde{k}_u)=(i_1,\ldots,i_u)$. 
This finishes the proof.
\end{IEEEproof}


Combining Proposition \ref{prop1} with Theorem \ref{thm1}, we bound below the redundancy of $(s,t)$-batch codes. 
\begin{theorem}\label{thm2}
\begin{enumerate}
    \item For $2\leq t\leq n$, 
    \begin{equation}\label{eq8}
        r(1,t)\geq \sqrt{2n+t^2-3t}-\frac{1}{2}.
    \end{equation}
    \item For $s\geq 2$ and $3s\leq t\leq n$,
    \begin{equation}\label{eq9}
        r(s,t)\geq 
           \sqrt{(s!)^{\frac{2}{s}}{n\choose s}^{\frac{1}{s}}+\frac{(\lfloor\frac{t}{s}\rfloor-3)^2}{4}}+\frac{\lfloor\frac{t}{s}\rfloor-1}{2}-s.
    \end{equation}
    Moreover, when $s\rightarrow\infty$ as $n\rightarrow\infty$, this turns to be
    \begin{equation}\label{eq10}
        r(s,t)\geq
           \left(\sqrt{\frac{sn}{e}+\frac{t^2}{4s^2}}+\frac{t}{2s}-s\right)\cdot(1-o(1)).
    \end{equation}
\end{enumerate}
\end{theorem}

\begin{IEEEproof}
For $s=1$, by Proposition \ref{prop1}, take $u=1$ and $v=t-1$ in (\ref{eq3}). We have  
\begin{equation}\label{eq10a}
    {{r+3-t+1}\choose 2}+(r+2-t+1)(t-3)\geq n,
\end{equation}
which leads to $r\geq \sqrt{2n+(t-\frac{5}{2})^2}-\frac{1}{2}$.

For $s\geq 2$ and $3s\leq t\leq n$, by Proposition \ref{prop1}, we can take $u=s$ and $v=\lfloor\frac{t}{s}\rfloor-1$ in Theorem \ref{thm1}. Note that ${{r+2u+1-v}\choose 2u}+\sum_{i=1}^{u}{{r+2u+1-(v+i)}\choose {2u-i}}{{v+i-3}\choose i}$ is the number of the monomials $x_1^{d_1}x_2^{d_2}\cdots x_{r}^{d_r}$ with $d_i$'s satisfying condition (\ref{eq6}). This number is upper bounded by the number of monomials $(x_1^{d_1}x_2^{d_2}\cdots x_{r}^{d_{r}})\cdot (y_{1}^{e_{1}}\cdots y_{r-v+2}^{e_{r-v+2}})$ of degree $2u$ with $d_i$ and $e_i$ satisfying
\begin{equation*}
\begin{cases}
   \sum_{i=1}^{r}d_i= u,\\
   \sum_{i=1}^{r-v+2}e_i= u,\\
   d_i\geq 0 \text{~for~} 1\leq i\leq r,\\
   e_i\geq 0 \text{~for~} 1\leq i\leq r.
\end{cases}
\end{equation*}
Therefore, we have 
\begin{equation*}
    {{r+u-1}\choose {u}}{{r+u+1-v}\choose u}\geq {{r+2u+1-v}\choose 2u}+\sum_{i=1}^{u}{{r+2u+1-(v+i)}\choose {2u-i}}{{v+i-3}\choose i}.
\end{equation*}
By substituting $u=s$ and $v=\lfloor\frac{t}{s}\rfloor-1$ into the above inequality, and using (\ref{eq3}), we obtain that 
\begin{align*}
    {{r+s-1}\choose {s}}{{r+s+2-\lfloor\frac{t}{s}\rfloor}\choose s} &\geq {n\choose s}.
\end{align*}
Note that ${a\choose b}\leq \frac{a^b}{b!}$ for every positive integers $a$ and $b$. Therefore,
\begin{align*}
    {{r+s-1}\choose {s}}{{r+s+2-\lfloor\frac{t}{s}\rfloor}\choose s} &\leq \frac{(r+s-1)^{s}(r+s+2-\lfloor\frac{t}{s}\rfloor)^{s}}{(s!)^2}\\
    &\leq \frac{1}{(s!)^2}\left((r+s-1)(r+s-1-(\lfloor\frac{t}{s}\rfloor-3))\right)^{s}\\
    &\leq \frac{1}{(s!)^2}\left((r+s-1)^2-(\lfloor\frac{t}{s}\rfloor-3)(r+s-1)\right)^{s}.
\end{align*} 
Thus, we have 
\[{n\choose s}\leq \frac{1}{(s!)^2}\left((r+s-1)^2-(\lfloor\frac{t}{s}\rfloor-3)(r+s-1)\right)^{s}.\]
This leads to 
\begin{align*}
\left((r+s-1)-\frac{\lfloor\frac{t}{s}\rfloor-3}{2}\right)^2\geq (s!)^{\frac{2}{s}}{n\choose s}^{\frac{1}{s}}+\frac{(\lfloor\frac{t}{s}\rfloor-3)^2}{4},
\end{align*}
which implies (\ref{eq9}). 
Equation (\ref{eq10}) follows from (\ref{eq9}) by Stirling's formula.
\end{IEEEproof}

\begin{remark}\label{rmk2}
\begin{itemize}
   \item Compared to the lower bound of $\sqrt{2n}$ on the redundancy $r(1,t)$ proved in \cite{RVW22}, Equation (\ref{eq8}) unifies the non-trivial lower bound $\sqrt{2n}$ when $t\leq \sqrt{2n}$ with the trivial lower bound $t$. 
   \item For $2\leq s\leq t/3$, Equation (\ref{eq9}) slightly improves upon the lower bound $r(s,t)\geq r(s,s)$. The improvement comes from
   \begin{equation*}
       {{r+2u+1-v}\choose 2u}+\sum_{i=1}^{u}{{r+2u+1-(v+i)}\choose {2u-i}}{{v+i-3}\choose i}<{{r+2u-1}\choose 2u}
   \end{equation*}
   when $v\geq 3$, where the LHS of the above inequality is the number of all monomials $x_1^{d_1}\cdots x_r^{d_r}$ satisfying condition (\ref{eq6}) and the RHS is the number of all monomials $x_1^{d_1}\cdots x_r^{d_r}$ of degree $2u$.
   \item The lower bound on $r(t,t)$ in \cite{AG21} follows from the lower bound on the redundancy of $\frac{t}{3}$-ordered-batch code (i.e., $(\frac{t}{3},2)$-ordered-batch code). In fact, in \cite{AG21} the authors showed that the redundancy $r$ of $(t,2)$-ordered-batch code satisfies
   \begin{equation*}
        {{r+2t-1}\choose 2t}\geq {n\choose t},
   \end{equation*}
   which is the case for $u=t$ and $v=2$ in (\ref{eq3}). However, there is a drawback of our proof: the field size in Theorem \ref{thm1} is required to be at least $poly(n)$.
\end{itemize}
\end{remark}

\section{Constructions of $(s,t)$-batch codes}
In this section, two constructions of $(s,t)$-batch codes are presented. 
The first construction is a recursive construction and the second is a random construction. 
For positive integers $u,v$, the recursive construction starts with a $(u,t)$-batch code and generate a $(uv,t)$-batch code. 
As a result, we obtain an upper bound on $r(s,t)$ using $r(1,t)$. 
The random construction follows a similar idea as in \cite{PPV20}. When applied on a binary $(s,t)$-batch code, we obtain that for $t=o(n^{\frac{1}{3}})$ and $s=O(\frac{t}{\log{n}})$, $r(s,t)\leq O(t^{\frac{3}{2}})\sqrt{n}$. 


\subsection{A recursive construction of $(s,t)$-batch code}
\label{sec:IV-rec}
We begin with the recursive construction. Before presenting the construction itself, we need some notations and definitions. 
For an integer $1\leq v\leq n$, a family of subsets $\mathcal{P}$ is called a \textbf{$v$-partition} of $[n]$ if $\mathcal{P}=\{P_1,\ldots,P_v\}$ is a partition of $[n]$ containing $v$ sets. For an integer $1\leq u\leq \frac{n}{v}$, a $uv$-subset $I=\{i_1,\ldots,i_{uv}\}\subseteq [n]$ is said to be \textbf{\emph{covered}} by a $v$-partition $\mathcal{P}=\{P_1,\ldots,P_v\}$ if for each $i\in [v]$, $|P_i\cap I|=u$. Clearly, if a $uv$-subset $I$ is covered by a $v$-partition $\mathcal{P}$, then $|P_i|\leq n-u(v-1)$ for each $i\in [v]$. Let $\mathfrak{P}=\{\mathcal{P}_1,\ldots, \mathcal{P}_S\}$ be a collection of $v$-partitions of $[n]$. $\mathfrak{P}$ is called \textbf{\emph{$u$-complete}} if every $uv$-subset of $[n]$ is covered by some $v$-partition $\mathcal{P}_i\in \mathfrak{P}$. 

Denote $S(n,u,v)$ as the minimum cardinality of a collection of $u$-complete $v$-partitions of $[n]$. Let $S=S(n,u,v)$ and $\mathfrak{P}=\{\mathcal{P}_1,\ldots, \mathcal{P}_S\}$ be a collection of $u$-complete $v$-partitions of $[n]$. Let $\mathcal{C}_0$ be a $(u,t)$-batch code of dimension $n_0=n-u(v-1)$ and redundancy $r_0=r(n_0;u,t)$ with encoder $\mathcal{E}_0$, where
\begin{equation*}
    \mathcal{E}_0: \mathbb{F}_q^{n_0}\to \mathbb{F}_q^{r_0} ,\; \mathbf{x}\mapsto \mathcal{E}_0(\mathbf{x})
\end{equation*}
such that $(\mathbf{x},\mathcal{E}_0(\mathbf{x}))\in\mathcal{C}_0$.
Based on $\mathfrak{P}$ and $\mathcal{C}_0$, we present the following construction.\\

\textbf{Construction I:} For every $\mathcal{P}_i\in\mathfrak{P}$, denote $\mathcal{P}_i=\{P_1^{i},\ldots,P_v^{i}\}$. Let $n\in \mathbb{N}$ be a positive integer and let $\mathbf{x}\in \mathbb{F}_q^{n}$ be an information vector. For every $j\in [v]$, $\mathbf{x}|_{P_j^{i}}$ is a vector over $\mathbb{F}_q$ of length $|P_j^{i}|\leq n_0$, obtained by considering the coordinates in $\mathbf{x}$ that appear in $P_j^i$. By appending $0$'s at the end of $\mathbf{x}|_{P_j^{i}}$, we can extend $\mathbf{x}|_{P_j^{i}}$ as $(\mathbf{x}|_{P_j^{i}},\mathbf{0})$ such that $(\mathbf{x}|_{P_j^{i}},\mathbf{0})\in\mathbb{F}_q^{n_0}$. Then, we define the encoder $\mathcal{E}$ as the following map 
\begin{align}\label{encoder}
    \mathcal{E}: \mathbf{x}\rightarrow (\mathcal{E}_0(\mathbf{x}|_{P_1^1},\mathbf{0}),\ldots,\mathcal{E}_0(\mathbf{x}|_{P_v^1},\mathbf{0}),\ldots,\mathcal{E}_0(\mathbf{x}|_{P_1^S},\mathbf{0}),\ldots,\mathcal{E}_0(\mathbf{x}|_{P_v^S},\mathbf{0})).
\end{align}
In other words, $\mathcal{E}(\mathbf{x})$ is the concatenation of vectors $\mathcal{E}_0(\mathbf{x}|_{P_j^i},\mathbf{0})$, $(i,j)\in [S]\times [v]$. Clearly, $\mathcal{E}$ is a map from $\mathbb{F}_q^{n}$ to $\mathbb{F}_q^{v\cdot S\cdot r_0}$. 
Then, we define the corresponding code $\mathcal{C}$ as 
$$\mathcal{C}=\{(\mathbf{x},\mathcal{E}(\mathbf{x})): \mathbf{x}\in \mathbb{F}_q^{n}\}\subseteq \mathbb{F}_q^{n+v\cdot S\cdot r_0}$$
and we call $\mathcal{C}$ the code defined by $\mathfrak{P}$ and $\mathcal{C}_0$.

\begin{theorem}\label{thm4}
Let $n,v,s,t$ be integers such that $1\leq v\leq s\leq t\leq n$, and denote $u=\lceil\frac{s}{v}\rceil$. Let $\mathfrak{P}=\{\mathcal{P}_1,\ldots, \mathcal{P}_S\}$ be a collection of $u$-complete $v$-partitions of $[n]$ with cardinality $S=S(n,u,v)$ and $\mathcal{C}_0$ be a $(u,t)$-batch code of dimension $n_0=n-u(v-1)$ and redundancy $r_0=r(n_0;u,t)$. Then, the code $\mathcal{C}$ defined by $\mathfrak{P}$ and $\mathcal{C}_0$ is an $(s,t)$-batch code of dimension $n$ and redundancy $v\cdot S \cdot r_0$. 
\end{theorem}

\begin{IEEEproof}
Let $\mathcal{E}_0$ be the encoder of $\mathcal{C}_0$.
By (\ref{encoder}) and the linearity of $\mathcal{E}_0$, $\mathcal{E}$ is a linear map. Thus, $\mathcal{C}$ is a systematic linear code of dimension $n$ and redundancy $v\cdot S \cdot r_0$. For every $(i,j)\in [S]\times [v]$, we denote $\mathbf{y}_{i,j}=\mathcal{E}_0(\mathbf{x}|_{P_j^{i}},\mathbf{0})$. Next, we assume that $s=uv$ and show that $\mathcal{C}$ is an $(s,t)$-batch code. The proof for the case when $s<uv$ is similar.

Let $I$ be a multi-set of requests of the form (\ref{eq_legalrequest}). W.l.o.g., assume that $I=\ms{1^{(a_1)},2^{(a_2)},\ldots, s^{(a_s)}}$ and $\sum_{i=1}^{s}a_{i}=t$. Since $\mathfrak{P}$ is $u$-complete, there is a $v$-partition $\mathcal{P}_l=\{P_1^{l},\ldots,P_v^{l}\}\in \mathfrak{P}$ such that $|P_j^{l}\cap \{1,\ldots,s\}|=u$ for every $j\in [v]$. Denote 
$$P_j^{l}\cap \{1,\ldots,s\}=\{i_{j,1},\ldots,i_{j,u}\}.$$ 
Then, we have $\sum_{h=1}^{u}a_{i_{j,h}}\leq t$. 
Recall that $(\mathbf{x}|_{P_j^{l}},\mathbf{0},\mathbf{y}_{l,j})\in \mathcal{C}_0$ is encoded by $\mathcal{E}_0$ from $(\mathbf{x}|_{P_j^{l}},\mathbf{0})$ and $\mathcal{C}_0$ is a $(u,t)$-batch code. Thus, requests $i_{j,1},\ldots,i_{j,u}$ can be supported with their multiplicities $a_{i_{j,1}},\ldots,a_{i_{j,u}}$ by accessing $\sum_{h=1}^{u}a_{i_{j,h}}$ mutually disjoint groups of symbols from $(\mathbf{x}|_{P_j^{l}},\mathbf{0},\mathbf{y}_{l,j})$. Note that symbols in the middle part of $(\mathbf{x}|_{P_j^{l}},\mathbf{0},\mathbf{y}_{l,j})$ are all $0$. Thus,  $i_{j,1},\ldots,i_{j,u}$ can actually be supported with their multiplicities by accessing $\sum_{h=1}^{u}a_{i_{j,h}}$ mutually disjoint groups of symbols from $(\mathbf{x}|_{P_j^{l}},\mathbf{y}_{l,j})$.

In general, since $\{1,\ldots,s\}= \bigcup_{j=1}^{v}\{i_{j,1},\ldots,i_{j,u}\}$, the analysis above implies that requests ${1,\ldots,s}$ can be supported with multiplicities $a_{1},\ldots,a_{s}$ by accessing $\sum_{h=1}^{u}a_{i_{j,h}}$ mutually disjoint groups of symbols from $(\mathbf{x}|_{P_j^{l}},\mathbf{y}_{l,j})$ for each $j\in [v]$. Moreover, since $P_1^{l},\ldots,P_v^{l}$ are mutually disjoint, $(\mathbf{x}|_{P_1^{l}},\mathbf{y}_{l,1}),\ldots,(\mathbf{x}|_{P_{v}^{l}},\mathbf{y}_{l,v})$ have disjoint support sets in $(\mathbf{x},\mathcal{E}(\mathbf{x}))$. Therefore, request $I$ can be supported with their multiplicities by accessing $t$ mutually disjoint groups of symbols from $(\mathbf{x},\mathcal{E}(\mathbf{x}))$. Therefore, we can conclude that $\mathcal{C}$ is an $(s,t)$-batch code.
\end{IEEEproof}

As an immediate corollary, we obtain 
%
\begin{corollary}\label{coro1}
For positive integers $n$ and $1\leq s\leq t\leq n$, the following holds
\begin{equation}\label{eq13}
    r(s,t)\leq s\left(\prod_{i=1}^{l}S(n,u_i,v_i)\right)r(1,t),
\end{equation}
where $u_i,v_i$ are integers satisfying $\prod_{i=1}^{l}v_i=s$ and $u_1=\frac{s}{v_1}$, $u_i=\frac{u_{i-1}}{v_i}$ for $2\leq i\leq l$.
\end{corollary}

\begin{IEEEproof} 
From Theorem \ref{thm4} we obtain the inequality 
\begin{equation}\label{eq12}
   r(n;uv,t)\leq v\cdot S(n,u,v)\cdot r(n-u(v-1);u,t),
\end{equation}
which implies that 
\begin{align}
    r(n;u_iv_i,t)&\leq v_iS(n,u_i,v_i)r(n-u_i(v_i-1);u_i,t) \nonumber\\
    \label{eq13b}&\leq v_iS(n,u_i,v_i)r(n;u_i,t)
\end{align}
for every $i\in [l]$. Note that $u_i=u_{i+1}v_{i+1}$. By applying (\ref{eq13b}) iteratively, this leads to 
\begin{align*}
    r(n;s,t)&\leq (\prod_{i=1}^{l}v_iS(n,u_i,v_i))r(n;u_l,t)\\
    &= s\left(\prod_{i=1}^{l}S(n,u_i,v_i)\right)r(n;1,t),
\end{align*}
where the last equality follows from $\prod_{i=1}^{l}v_i=s$ and $u_l=\frac{s}{\prod_{i=1}^{l}v_i}$.
\end{IEEEproof}

In order to use Corollary \ref{coro1} to estimate $r(s,t)$, we bound above $S(n,u,v)$.

\begin{lemma}\label{lem1}
For positive integers $n$, $u$ and $v$ such that $u\leq \frac{n}{v}$, the following holds
\begin{equation}\label{eq14}
    S(n,u,v)\leq \big\lfloor\frac{\ln{n\choose uv}}{-\ln\left(1-\prod_{i=1}^{v}\frac{{{(v-i+1)u}\choose u}}{v^{u}}\right)}\big\rfloor+1.
\end{equation}
\end{lemma}

\begin{IEEEproof}
For a $v$-partition $\mathcal{P}=\{P_1,\ldots,P_v\}$ of $[n]$, we define the vector $\mathbf{x}_{\mathcal{P}}\in [v]^n$: for every $i\in [n]$, $\mathbf{x}_{\mathcal{P}}(i)=j$ if $i\in P_j$. Meanwhile, every vector in $[v]^{n}$ can represent a $v$-partition of $[n]$ if we allow some parts of the partition to be empty. Let $I=\{i_1,\ldots,i_{uv}\}$ be a $uv$-subset of $[n]$. 
Then, $I$ is covered by $\mathcal{P}$ if and only if $|P_i\cap I|=u$ for every $i\in [v]$, i.e., the projection of $\mathbf{x}_{\mathcal{P}}$ on indexes $\{i_1,\ldots,i_{uv}\}$ forms a permutation of
\begin{equation}\label{eq14a}
(\underbrace{1,\ldots,1}_{u},\underbrace{2,\ldots,2}_{u},\ldots, \underbrace{v,\ldots,v}_{u}).
\end{equation}
For simplicity, we say that $I$ is covered by a vector $\mathbf{x}\in[v]^{n}$ if $I$ is covered by the $v$-partition of $[n]$ corresponding to $\mathbf{x}$.

For a fixed $uv$-subset $I\subseteq [n]$, choose vector $\mathbf{x}$ from $[v]^{n}$ uniformly at random. Then, the probability that $\mathbf{x}$ covers $I$ is 
$$\text{Pr}(\mathbf{x}|_{I}\text{~forms a permutation as in (\ref{eq14a})})=\prod_{i=1}^{v}\frac{{{(v-i+1)u}\choose u}}{v^{u}}.$$
Hence, if we draw some $S$ vectors independently uniformly at random from $[v]^{n}$, then the probability that none of the $v$-partitions corresponding to these $S$ vectors cover $I$ is 
$$\left(1-\prod_{i=1}^{v}\frac{{{(v-i+1)u}\choose u}}{v^{u}}\right)^{S}.$$
For each $uv$-subset $I\subseteq [n]$, let $Y_{I}$ be the corresponding indicator random variable:
\begin{equation*}
    Y_{I}=\begin{cases}
    0,~\text{if $I$ is covered by at least one of the $S$ vectors};\\
    1,~\text{otherwise}. 
    \end{cases}
\end{equation*}
Denote $Z=\sum_{I\subseteq {[n]\choose uv}}Y_{I}$.
By the linearity of expectation, we have
$$\mathbb{E}[Z]=\sum_{I\subseteq {[n]\choose uv}}\mathbb{E}[Y_I]={n\choose uv}\left(1-\prod_{i=1}^{v}\frac{{{(v-i+1)u}\choose u}}{v^{u}}\right)^{S}.$$

Using Markov's inequality we obtain that $\Pr(Z=0)\geq 1-\mathbb{E}[Z]$. 
Therefore, if $\mathbb{E}[Z]< 1$, then there exists a selection of $S$ vectors from $[v]^{n}$ for which $Z=0$. This means that all the $uv$-subsets of $[n]$ will be covered by at least one of these $S$ vectors. When 
$$S=\left\lfloor\frac{\ln{n\choose uv}}{-\ln\left(1-\prod_{i=1}^{v}\frac{{{(v-i+1)u}\choose u}}{v^{u}}\right)}\right\rfloor+1,$$ 
we have
\begin{align*}
    \left(1-\prod_{i=1}^{v}\frac{{{(v-i+1)u}\choose u}}{v^{u}}\right)^{S}&=e^{\ln\left(1-\prod_{i=1}^{v}\frac{{{(v-i+1)u}\choose u}}{v^{u}}\right)S}\\
    &< e^{-\ln{n\choose uv}},
\end{align*}
which leads to $\mathbb{E}[Z]<1$. 
\end{IEEEproof}


\begin{corollary}\label{coro2}
For $n$ large enough and $1\leq s\leq t=o(n)$, the following holds. 
\begin{itemize}
    \item [1.] When $s$ is a constant, we have
    \begin{equation}\label{eq15}
    r(s,t)\leq O(\ln{n})\cdot r(1,t).
    \end{equation}
    \item [2.] When $s\rightarrow\infty$ as $n\rightarrow\infty$, we have
    \begin{equation}\label{eq16a}
    r(s,t)\leq O(s^{1.5}e^s\ln{n})\cdot r(1,t).
    \end{equation}
    Moreover, if $s$ also satisfies $2^{l-1}<s\leq 2^{l}$ for some $l=o(\ln(n))$, then (\ref{eq16a}) can be modified as
    \begin{equation}\label{eq16b}
    r(s,t)\leq O(2^{\frac{3l^2+l}{4}}(\ln{n})^{l})\cdot r(1,t).
    \end{equation}
\end{itemize}
\end{corollary}

\begin{IEEEproof}
Let $u=1$ and $v=s$ in Theorem \ref{thm4}, using \eqref{eq12} we have 
$$r(s,t)\leq s\cdot S(n,1,s)\cdot r(1,t).$$
By Lemma \ref{lem1}, we have
$$S(n,1,s)\leq \left\lfloor\frac{\ln{n\choose s}}{-\ln\left(1-\prod_{i=1}^{s}\frac{s-i+1}{s}\right)}\right\rfloor+1$$
and (\ref{eq15}) follows from the assumption that $s$ is a constant.

When $s\rightarrow\infty$ as $n\rightarrow\infty$, by Stirling formula and since $\ln{(1-x)}=-x(1+o(1))$ as $x\rightarrow 0$, we have  
\begin{align*}
S(n,1,s)&\leq\left\lfloor\frac{\ln{n\choose s}}{-\ln\left(1-\frac{s!}{s^{s}}\right)}\right\rfloor+1 \leq O\left(\frac{s\ln{n}}{s!/s^s}\right)\\
    &\leq O\left(\frac{s\ln{n}}{\sqrt{s}/e^s}\right)=O(s^{0.5}e^s\ln{n}).
\end{align*}
Let $u=1$ and $v=s$ in Theorem \ref{thm4}, this implies (\ref{eq16a}). Specially, when $s=2^l$, by Corollary \ref{coro1}, 
$$r(2^{l},t)\leq s\left(\prod_{i=1}^{l}S(n,2^{l-i},2)\right)r(1,t).$$
Again, by Stirling's formula and $\ln{(1-x)}=-x(1+o(1))$ as $x\rightarrow 0$, we have
\begin{align*}
    S(n,u,2)&\leq \left\lfloor\frac{\ln{n\choose 2u}}{-\ln\left(1-\prod_{i=1}^{2}\frac{{{(2-i+1)u}\choose u}}{2^{u}}\right)}\right\rfloor+1\\
    &=O\left(\frac{u\ln{n}}{{2u\choose u}/{2^{2u}}}\right)= O\left(u^{\frac{3}{2}}\ln{n}\right).
\end{align*}
Therefore,
\begin{align*}
    r(2^{l},t)&\leq 2^{l}\cdot O\left(\prod_{i=1}^{l}\parenv{2^{\frac{3(l-i)}{2}}\ln{n}}\right)\cdot r(1,t)\\
    &\leq O\left(2^{\frac{3l^2+l}{4}}(\ln{n})^{l}\right)\cdot r(1,t),
\end{align*}
where the last inequality follows from $\sum_{i=1}^{l}\frac{3(l-i)}{2}+l=\frac{3(l^2-l)}{4}+l=\frac{3l^2+l}{4}$. Then, (\ref{eq16b}) follows from $r(s,t)\leq r(2^{l},t)$.
\end{IEEEproof}

\begin{remark}\label{rmk4}
Construction I can be viewed as a generalization of the construction of $(t,t)$-batch code that appears in \cite{VY16}.
Using a construction of $\lfloor\frac{t}{2}\rfloor$-complete $\lfloor\frac{t}{2}\rfloor$-partitions, the authors of \cite{VY16} obtained an upper bound on $r(t,t)$ by $r(1,t)$ and showed that $r(t,t)=O(\sqrt{n}\ln{n})$ for constant $t$. 
\end{remark}

\subsection{A random construction of $(s,t)$-batch codes}\label{sec_IVb}

Through a random construction based on point-line incidences on the plane $\mathbb{F}_q^2$, Polyanskaya et al. \cite{PPV20} showed that $r(t,t)=O(t^{\frac{3}{2}}\sqrt{n}\ln{n})$ for $t=O(\frac{n^{\frac{1}{3}}}{\ln {n}})$. In this subsection, we modify their random construction and prove the following result.


\begin{theorem}\label{thm5}
For $n$ large enough, let $t$ be an integer such that $8\ln{n}\leq t\leq n^{\frac{1}{3}}$. Then, for all $1\leq s \leq \frac{t}{8\ln{n}}$, there exists a binary $(s,t)$-batch code of dimension $n$ with at most $O(t^{\frac{3}{2}}n^{\frac{1}{2}})$ redundancy symbols. Thus, 
\begin{equation}\label{eq17}
    r(s,t)\leq O(t^{\frac{3}{2}}\sqrt{n}).
\end{equation}
\end{theorem}

Before proving Theorem \ref{thm5}, we need some notations to present the construction of the code. Let $q$ be a prime power. 
Let $(P,\mathcal{L})$ denote the finite affine plane of order $q$, where $P$ is the set of $n=q^2$ points and $\mathcal{L}$ is the set of all the $n+q$ lines over the plane. Then, any two lines in $\mathcal{L}$ intersect in at most one point, each line in $\mathcal{L}$ contains $q$ points and each point in $P$ is contained in $q+1$ lines. \\

\textbf{Construction II}: 
\begin{enumerate}
    \item Let $q$ be a prime power and set $n=q^2$. Let $(P,\mathcal{L})$ denote the finite affine plane of order $q$. 
    \item Let $0< p_1,p_2<1$ be two parameters to be determined later. 
To start with, we construct a family $\mathcal{F}$ of lines in $\mathcal{L}$ as follows: 
\begin{enumerate}
    \item[\underline{Step 1:}] Choose each line from $\mathcal{L}$ independently with probability $p_1$ (without repetition), and denote the resulting set of lines as $\mathcal{F}$. 
    
    \item[\underline{Step 2:}] We operate on each line $L\in \mathcal{L}$ independently and construct a subset of points $R(L)\subseteq L$ by picking each point on $L$ independently with probability $p_2$.
\end{enumerate} 

\item Construct a systematic linear code $\mathcal{C}$ of dimension $n$ and redundancy at most $|\mathcal{F}|$ using $(P,\mathcal{L})$ and $\mathcal{F}$ as follows. Fix a bijection that associates $n$ information symbols with points in $P$. Next, for every $L\in\mathcal{F}$, if $R(L)\neq\emptyset$, define a parity-check symbol $\mathbf{y}_{L}=\sum_{\mathbf{x}\in R(L)}\mathbf{x}$. 
\end{enumerate} 
With a slight abuse of notation, we use points $\mathbf{x}\in P$ to represent the index of $\mathbf{x}$ in $[n]$. 

Notice that Construction II may fail to yield an $(s,t)$-batch code with the mentioned redundancy. 
The construction fails if there exists a multi-set of requests $I$ for which there are no mutually disjoint recovering sets. 
Therefore, to prove the theorem we aim to demonstrate that with high probability, the code $\cC$ resulting from Construction II, has the following property: 
For any multi-set of requests $I=\ms{{i_1}^{(a_1)},{i_2}^{(a_2)},\ldots, {i_s}^{(a_{s})}}\subseteq [n]$ such that $\sum_{l=1}^{s}a_l=t$, 
there exist $a_l$ distinct recovering sets for each $i_l$. 
Furthermore, these $t$ recovering sets are pairwise disjoint. 
It is noteworthy that the value of $a_j$ cannot surpass $q+1$, given that each $x_i$ is contained in at most $q+1$ lines. 
Indeed, $a_j\leq t$ and our assumption holds that $t\leq n^{\frac{1}{3}}$. 
Since $n=q^2$, it follows that $t\leq q^{2/3}\leq q+1$. 

To accomplish this objective, we employ an algorithm that given a set of requests, finds mutually disjoint recovering sets with high probability. 
Let $I$ be a multi-set 
\[I=\ms{{i_1}^{(a_1)},{i_2}^{(a_2)},\ldots, {i_s}^{(a_{s})}}\] 
with $\sum_{l=1}^{s}a_l=t$. 
Assume that for a certain $1\leq j\leq s$, we have already established $a_l, \; 1\leq l\leq j-1$ recovering sets for $i_l$. 
Moreover, assume that all these $\sum_{l=1}^{j-1}a_l$ recovering sets are mutually disjoint. 

To proceed with finding the subsequent $a_j$ recovering sets, consider all the lines in $\mathcal{F}$ that pass through $\mathbf{x}_{i_j}$. 
Find $a_j$ distinct lines $L_{j,1},\ldots,L_{j,a_j}$ such that for $1\leq l\leq a_j$, we have  
$L_{j,l}\cap \left(\{\mathbf{x}_{i_1},\ldots,\mathbf{x}_{i_s}\}\setminus\{\mathbf{x}_{i_{j}}\}\right)=\emptyset$, and $R(L_{j,l})$ contains 
$\mathbf{x}_{i_j}$. 
Define $\{\mathbf{y}_{R_{j,1}}\}\cup R(L_{j,1})\setminus\{\mathbf{x}_{i_j}\},\ldots,\{\mathbf{y}_{R_{j,a_j}}\}\cup R(L_{j,a_j})\setminus\{\mathbf{x}_{i_j}\}$ as the recovering sets, where $\mathbf{y}_{R_{j,l}}$ is the parity-check symbol defined in Construction II. 
Notice that since $L_{j,1},\ldots,L_{j,{a_j}}$ are different lines that contain $\bbx_{i_j}$, the recovering sets 
$\{\mathbf{y}_{R_{j,1}}\}\cup R(L_{j,1})\setminus\{\mathbf{x}_{i_j}\},\ldots,\{\mathbf{y}_{R_{j,a_j}}\}\cup R(L_{j,{a_j}})\setminus\{\mathbf{x}_{i_j}\}$ 
are mutually disjoint. 
Repeat this process until $j=s$, i.e., until all the recovering sets are obtained.\\ 

To prove the theorem, we need to estimate the probability that there exists a multi-set $I$ for which the algorithm above fails. 
Let us simplify notations by denoting $R_{j,l}:=R(L_{j,l})$ and $\mathbf{y}_{j,l}:=\mathbf{y}_{R_{j,l}}$. 
For every multi-set $I$ and every round $j$, the construction fails if at least one of the following occurs. 
 \begin{enumerate} 
    \item There are no $a_j$ distinct lines $L_{j,1},\ldots,L_{j,a_j}$ that contain $\mathbf{x}_{i_j}$. 
    \item There exists $1\leq l\leq a_j$ for which $L_{j,l}$ goes through a point in 
            $\{\mathbf{x}_{i_1},\ldots,\mathbf{x}_{i_s}\}\setminus\{\mathbf{x}_{i_j}\}$. 
    \item The (mutually disjoint) recovering sets 
            $\{\mathbf{y}_{j,1}\}\cup R_{j,1}\setminus\{\mathbf{x}_{i_j}\},\ldots, \{\mathbf{y}_{j,a_j}\} \cup R_{j,a_j}\setminus\{\mathbf{x}_{i_j}\}$ are not mutually disjoint with the previously chosen recovering sets. 
    \item The redundancy of the code is larger than $O(t^{1/3}\sqrt{n})$, i.e., $|\cF|>O(t^{1/3}\sqrt{n})$.
 \end{enumerate}

To estimate the size $|\cF|$, we notice that it is binomially distributed $|\mathcal{F}|\sim B(n+q, p_1)$ and we use concentration bounds. 
To estimate the probability of the rest of the failure points, we use concentration bounds together with the fact that for every $L\in \mathcal{L}$, 
$|R(L)|\sim B(q, p_2)$, which implies that 
$$\abs{\{L:~L\in \mathcal{F}~\text{and}~\mathbf{x}\in R(L)\}}\sim B(q+1, p_1p_2).$$

To simplify the proof of Theorem \ref{thm5}, we use several lemmas. 
The first lemma is the Chernoff bound. The following version is sometimes referred to as the multiplicative Chernoff bound. 
Since we could only find a slightly different version of this bound (see \cite[Thm. 4.4 and Thm. 4.5]{Prob2017}), we added a short proof of this version in the appendix for completeness . 
\begin{lemma}[Chernoff bound] 
\label{lem:chern}
Let $X=\sum_{i=1}^{n}X_i$, where $X_i=1$ with probability $p_i$ and $X_i=0$ with
probability $1-p_i$, and all $X_i$ are independent. Let $\mu=\E[X]$. Then, 
\[\Pr(X\geq (1+\delta)\mu)\leq e^{-\frac{\delta^2\mu}{2+\delta}},\qquad \forall~ \delta>0,\] 
and 
\[\Pr(X\leq (1-\delta)\mu)\leq e^{-\mu\delta^2/2},\qquad \forall~ 0<\delta<1.\]
\end{lemma}

We now notice that in Construction II, we choose $\cF$ randomly. Then, the set $\cF$ is used to produce the redundancy bits of the constructed code $\cC$. 
To show that the redundancy of the code is small, we first bound the size $|\cF|$. 
\begin{lemma}
    \label{lem:help1} 
    Let $(P,\mathcal{L})$ and $\mathcal{F}$ be chosen according to Construction II. Let $0<p_1,p_2<1$ be two parameters to be determined later. 
    Then 
    \[\Pr\parenv{|\cF|\geq 3p_1 n}\leq e^{-\frac{p_1(n+q)}{3}}.\]
\end{lemma}

\begin{IEEEproof}
    We notice that every line from $\cL$ is chosen to $\cF$ independently with probability $p_1$. 
    Thus, the size $|\cF|$ is binomially distributed $B(n+q,p_1)$. 
    This implies that $\E\sparenv{|\cF|}=p_1(n+q)$, and since $q\geq 2$ and $n=q^2\geq 2q$, we have $3p_1n\geq 2p_1(n+q)$. 
    Putting everything together and using Chernoff bound with $\delta=1$ we obtain 
    \begin{align*}
        \Pr\parenv{|\cF|\geq 3p_1n} &\leq \Pr\parenv{|\cF|\geq 2p_1n}\\ 
        &\leq e^{-\frac{p_1(n+q)}{3}}.
    \end{align*}
\end{IEEEproof}

The next lemma bounds the probability that Construction II fails to generate an $(s,t)$-batch code. 
\begin{lemma}
    \label{lem:help2} 
    Consider Construction II with the parameters $p_2=\frac{1}{2\sqrt{2t}}$ and $p_1=24\sqrt{2}\frac{t^{3/2}}{q}$. 
    Assume also that $t$ is an integer such that $8\ln n\leq t\leq n^{1/3}$ and $1\leq s\leq \frac{t}{8\ln n}$. 
    Let $A$ denote the event that Construction II fails to generate an $(s,t)$-batch code. Then 
    \[\Pr(A)\leq 2ne^{-\frac{n^{1/3}}{6\sqrt{2}}} + e^{-t/4}.\]
\end{lemma}

\begin{IEEEproof}
    We begin the proof by recalling that Construction II fails to generate an $(s,t)$-batch code if there exists a multi-set of requests 
    $I$ for which there are no mutually disjoint recovering sets. 
    For $i\in [n]$ and for $0\leq w<t$, we define $A_{i,w}$ to be the event that there exists a multi-set 
    $I=\ms{{i_1}^{(a_1)},{i_2}^{(a_2)},\ldots, {i_s}^{(a_{s})}}$ with $i_j=i$ for some $j\in [s]$ and $a_j=t-w$, such that the above algorithm 
    finds $\sum_{l=1}^{j-1}a_l$ mutually disjoint recovering sets for $i_1,\dots,i_{j-1}$, but fails to find $a_j$ recovering sets for $i_j$. 
    Let $C$ denote the event that there is a line $L\in\cF$ such that $|R(L)|>2p_2q$. 
    We have that 
    \begin{align}\label{eq21}
        \Pr(A)&\leq \Pr\parenv{\bigcup_{i\in [n] \atop 0\leq w<t} A_{i,w}}\nonumber\\ 
        &\leq \Pr(C)+\Pr\parenv{\bigcup_{i\in [n] \atop 0\leq w<t} (A_{i,w}\cap \overline{C})}\nonumber\\ 
        &\leq \Pr(C)+tn\cdot\max_{i\in [n] \atop 0\leq w<t}\Pr(A_{i,w}\cap \overline{C}).
    \end{align}

    We first estimate $\Pr(C)$. For a fixed line $L$, every point in $R(L)$ is chosen independently with probability $p_2$. 
    Thus, the size $|R(L)|$ is distributed binomially $B(q,p_2)$ with $\E[|R(L)|]=p_2q$. 
    By Chernoff bound (with $\delta=1$) we obtain that for every line $L\in \cF$, 
    \[\Pr\parenv{|R(L)|\geq 2p_2q}\leq e^{-\frac{p_2q}{3}}.\] 
    Plugging $p_2$ into $p_2q$ we get $p_2q=\frac{q}{2\sqrt{2t}}=\frac{\sqrt{n}}{2\sqrt{2t}}$. 
    Using the assumption that $t\leq n^{\frac{1}{3}}$ we have $p_2q\geq \frac{n^{1/3}}{2\sqrt{2}}$. 
    This implies that 
    \[\Pr(C)\leq 2ne^{-\frac{n^{1/3}}{6\sqrt{2}}}.\] 

    We now estimate the probability $\Pr(A_{i,w}\cap \overline{C})$. 
    To that end, for every $i\in [n]$, denote by $C_i$ the event that every line $L\in\cF$ that does not contain $\bbx_i$, contains at most   
    $2p_2q$ points in $R(L)$, i.e., $|R(L)|\leq 2p_2q$. 
    Notice that $\overline{C_i}$ is the event that there is a line $L\in\cF$ with $|R(L)|>2p_2q$ and $\bbx_i\notin L$. 
    It is immediate that $\overline{C_i}\subseteq C$, which implies 
    \[\Pr(A_{i,w}\cap \overline{C})\leq \Pr(A_{i,w}\cap C_i).\] 

    For a multi-set $I=\ms{{i_1}^{(a_1)},{i_2}^{(a_2)},\ldots, {i_s}^{(a_{s})}}$ with $i_j=i$ for some $j\in [s]$, let $D_1(I)$ denote the event that the algorithm fails to find $a_{j}$ desired recovering sets for $i$. 
    Denote by $D_2(I)$ the event that the algorithm manages to find $\sum_{l=1}^{j-1} a_{l}$ desired recovering sets for $i_1,\dots, i_{j-1}$. 
    From the definition of $A_{i,w}$ and the union bound, we have 
    \[\Pr(A_{i,w}\cap C_i)\leq \sum_{I=\ms{{i_1}^{(a_1)},\ldots, {i_s}^{(a_{s})}} \atop i_j=i \text{~and~} w=t-a_j}\Pr(D_1(I)\cap D_2(I)\cap C_i).\]
    The number of possible multi-sets containing $i$ with multiplicity $t-w$ is at most ${n\choose s-1}{w+s-1\choose s-1}$, since we first select the rest of the coordinates, and then choosing multiplicities for every coordinate. 
    Therefore, we have 
    \begin{align}\label{eq22a}
    \Pr(A_{i,w}\cap C_i)&\leq {n\choose s-1}{w+s-1\choose s-1}\cdot \Pr(D_1(I)\cap D_2(I)\cap C_i)\nonumber\\
    &={n\choose s-1}{w+s-1\choose s-1}\cdot \Pr(D_1(I) | D_2(I)\cap C_i)\cdot \Pr(D_2(I)\cap C_i)\nonumber\\
    &\leq {n\choose s-1}{t+s-1\choose s-1}\cdot \Pr(D_1(I) | D_2(I)\cap C_i). 
    \end{align}

    We are now left to estimate $\Pr(D_1(I) | D_2(I)\cap C_i)$ for a multi-set $I$. 
    To that end, fix a multi-set $I$, and assume w.l.o.g. that $1\leq i\leq s$ and $I=\ms{1^{(a_1)},2^{(a_2)},\ldots, s^{(a_{s})}}$.
    By the definition of $C_i$, if $L$ does not contain $\mathbf{x}_i$, then $|R(L)|< 2qp_2$. 
    Moreover, the event $D_2(I)$ implies that for every $1\leq j<i$ and $1\leq l\leq a_j$, line $L_{j,l}$ does not contain $\mathbf{x}_i$. 
    Therefore, 
    \begin{equation}\label{eq22b}
        \abs{\bigcup_{j=1}^{i-1}\bigcup_{l=1}^{a_j}R_{j,l}}\leq 
        \left(\sum_{j<i}a_j\right)\cdot 2p_2q\leq 2tp_2q. 
    \end{equation}
    Denote $\{L_{1},\ldots, L_{q+1}\}\subseteq \mathcal{L}$ as the set of lines passing through $\mathbf{x}_i$. 
    By (\ref{eq22b}), we have 
    $$\abs{\left(\bigcup_{l=1}^{q+1}L_l\setminus \{\mathbf{x}_i\}\right)\cap \left(\bigcup_{j=1}^{i-1} \bigcup_{l=1}^{a_j} R_{j,l}\right)}\leq 2tp_2q.$$ 
    Recall that two distinct lines intersect at most at one point. 
    Since the lines $L_1,\dots, L_{q+1}$ are distinct and contain $\bbx_i$, there are no other common points. 
    This implies that the number of lines in $\mathset{L_1,\dots, L_{q+1}}$ that intersect $\bigcup_{j=1}^{i-1}\bigcup_{l=1}^{a_{j}}R_{j,l}$ with at least $4p_2t$ points, is at most $q/2$. 
    In other words, 
    \[\abs{\mathset{i ~:~ 1\leq i\leq q+1,\; \abs{L_i\cap \parenv{\bigcup_{j=1}^{i-1}\bigcup_{l=1}^{a_j}R_{j,l}}}\geq 4p_2t}}\leq \frac{q}{2}.\] 
    Additionally, there are at most $s-1$ lines from $\mathset{L_1,\dots, L_{q+1}}$ passing through some points in $\mathset{\bbx_1,\dots,\bbx_s}\setminus \mathset{\bbx_i}$. 
    For $q$ large enough, we have $s\leq q/6$, and thus 
    \[\abs{\mathset{i ~:~ 1\leq i\leq q+1,\; L_i\cap \mathset{\bbx_1,\dots,\bbx_s}\setminus \mathset{\bbx_i}=\emptyset,\; \abs{L_i\cap \bigcup_{j=1}^{i-1}\bigcup_{l=1}^{a_j}R_{j,l}}\leq 4p_2t }}\geq q+1-\frac{q}{2}-(s-1)\geq \frac{q}{3}.\] 
    In words, there are at least $q/3$ lines in $\mathset{L_1,\dots, L_{q+1}}$ that do not contain any point from $\mathset{\bbx_1,\dots,\bbx_s}\setminus \mathset{\bbx_i}$, and intersect $\bigcup_{j=1}^{i-1}\bigcup_{l=1}^{a_j}R_{j,l}$ with less than $4p_2t$ points. Let us denote $h=\floorenv{q+1-q/2-(s-1)}$. 
    W.l.o.g., assume that these lines are $\{L_{1},\ldots, L_h\}$.\\  

    Let $\zeta_{1},\ldots,\zeta_{h}$ be binary random variables such that $\zeta_{j}=1$ if and only if $L_j\in \cF$, $\bbx_i\in R(L_j)$, and 
    $R(L_{j})\cap \parenv{\bigcup_{j=1}^{i-1}\bigcup_{l=1}^{a_j}R_{j,l}}=\emptyset$. 
    Notice that $\zeta_j$ are independent random variables since each line is chosen independently, and each point from a chosen line is also chosen independently. 
    We have that 
    \[\Pr(\zeta_j=1)\geq p_1p_2(1-p_2)^{4p_2t}.\] 
    Indeed, $p_1$ is the probability that $L_j$ was selected to $\cF$, $p_2$ is the probability that $\bbx_i$ is to $R(L_j)$, and since $L_1,\dots, L_h$  intersect 
    $\bigcup_{j=1}^{i-1}\bigcup_{l=1}^{a_j}R_{j,l}$ with less then $4p_2t$ points, $(1-p_2)^{4p_2t}$ is the probability that no points from $ \bigcup_{j=1}^{i-1}\bigcup_{l=1}^{a_j}R_{j,l}$ were selected. 

    Let $w$ be such that $a_i=t-w$. If $\sum_{j=1}^{h}\zeta_j\geq t-w$ then the construction of $a_j$ recovering sets succeeds. Thus 
    \[\Pr(D_1(I) ~|~ D_2(I)\cap C_j)\leq \Pr(\sum_{j=1}^h \zeta_j <t-w).\] 
    From Chernoff bound (Lemma \ref{lem:chern}) we obtain 
    \begin{align}\label{eq23}
        \Pr(\sum_{j=1}^h \zeta_j <t-w)&\leq e^{-\frac{\sum_{j=1}^{\frac{q}{3}}\Pr(\zeta_j=1)}{2}\cdot\delta^2}\nonumber\\ 
        &\overset{(a)}{\leq} e^{-\frac{qp_1p_2(1-p_2)^{4p_2t}}{6}\cdot\delta^2}, 
    \end{align} 
    where $\delta$ satisfies $(1-\delta)\sum_{j=1}^{\frac{q}{3}}\Pr(\zeta_j=1)=t-w$ and $(a)$ follows since $\Pr(\zeta_j=1)\geq p_1p_2(1-p_2)^{4p_2t}$. 
    Using the inequality $(1-p)^x\geq 1-px$ which holds for $0\leq p\leq 1$ and $x\geq 0$, we obtain 
    \[(1-p_2)^{4p_2t}\geq 1-4p_2^2 t.\] 
    Plugging in $p_2=\frac{1}{2\sqrt{2t}}$ we obtain 
    \[1-4p_2^2t\geq \frac{1}{2}.\] 
    Plugging in $p_1=24\sqrt{2}\frac{t^{3/2}}{q}$, we have 
    \[qp_1p_2(1-p_2)^{4p_2t}\geq \frac{qp_1p_2}{2}\geq 6t.\]
    Thus, we obtain that for $t\geq 2$, 
    \begin{align*} 
        \delta=1-\frac{t-w}{\sum_{j=1}^{\frac{q}{3}}\Pr(\zeta_j=1)}&\geq 1-\frac{t}{\sum_{j=1}^{\frac{q}{3}}\Pr(\zeta_j=1)}\\ 
        &\geq 1-\frac{3}{qp_1p_2(1-p_2)^{4p_2t}}\\
        &\geq 1-\frac{1}{2t}\geq\frac{1}{\sqrt{2}}.
    \end{align*}
    Putting everything together we obtain 
    \[\Pr(D_1(I) ~|~ D_2(I)\cap C_j)\leq e^{-t\delta^2}\leq e^{-t/2}.\]

    Substituting this upper bound in (\ref{eq22a}), we obtain 
    \begin{align*}
        tn\cdot\max_{i\in [n] \atop 0\leq w<t}\Pr(A_{i,w}\cap \overline{C})& \leq tn\cdot n^{s-1}\cdot (t+s-1)^{s-1}\cdot e^{-\frac{t}{2}}\\ 
        &\leq n^{s}\cdot (t+s-1)^{s}\cdot e^{-\frac{t}{2}}\\ 
        &\overset{(a)}{\leq} e^{2s\ln{n}-\frac{t}{2}}\overset{(b)}{\leq} e^{-\frac{t}{4}},
    \end{align*} 
    where $(a)$ follows from $t+s-1\leq n$ and $(b)$ follows from $s\leq \frac{t}{8\ln{n}}$. 

    Putting everything together, we get 
    $$\Pr(A)\leq 2ne^{-\frac{n^{1/3}}{6\sqrt{2}}}+e^{-\frac{t}{4}},$$ 
    which proves the lemma.
\end{IEEEproof}

The proof of Theorem \ref{thm5} now follows. 
\begin{IEEEproof}[Proof of Theorem \ref{thm5}]
We notice first that Construction II fails if either the constructed code is not an $(s,t)$-batch code or if the redundancy of the code is larger than $O(t^{1/3}\sqrt{n})$. 
Let us denote by $A$ the event that Construction II fails to construct an $(s,t)$-batch code, and by $B$ the event that the redundancy of the code is larger than $3p_1n$. 
Plugging in $p_1=24\sqrt{2}\frac{\sqrt{t^3}}{q}$ yields 
$3p_1n=O(t^{\frac{3}{2}}\sqrt{n})$. The event $B$ is the event that $|\cF|\geq 3p_1n$. 
Thus, the probability that Construction II fails is given by 
\[\Pr(A\cup B)\leq \Pr(A)+\Pr(B).\] 

From Lemma \ref{lem:help1}, $\Pr(B)\leq e^{-\frac{p_1(n+q)}{3}}$. 
Plugging in $p_1=24\sqrt{2}\frac{\sqrt{t^3}}{q}$ yields 
\[\Pr(B)\leq e^{-\frac{\sqrt{n}}{3}}.\] 

Together with Lemma \ref{lem:help2} we obtain 
\[\Pr(A\cup B)\leq e^{-\frac{\sqrt{n}}{3}}+2ne^{-\frac{n^{1/3}}{6\sqrt{2}}}+e^{-\frac{t}{4}}.\] 
Since for large enough $n$, this probability is strictly less than $1$, such a code exists and the theorem follows.
\end{IEEEproof}

\section{Functional $(s,t)$-batch codes}

As another generalization of PIR codes and batch codes, \emph{functional batch codes} was introduced by Zhang, Etzion and Yaakobi \cite{ZEY20}. Instead of just dealing with the multi-set requests of information symbols, functional batch codes can support parallel requests of linear combinations of information symbols. In previous sections, we studied the trade-off between the $(s,t)$-batch code property and the minimum required redundancy. In this section, we study this trade-off for functional $(s,t)$-batch codes.

\subsection{Definitions and related results}

We start by introducing the definition of a \emph{functional $(s,t)$-batch code}. 
\begin{definition}\label{GBFC}
A (binary) \emph{functional $(s,t)$-batch code} $\cC_F(s,t)$ of length $N$ and dimension $n$ is a systematic linear code that encodes $n$ information symbols $x_1,\ldots,x_{n}$. For each $1\leq i\leq N$, the $i$-th code symbol, $c_i$, is a non-trivial linear combination of $x_1,\ldots,x_{n}$. 
Let $\mathbf{v}\in \{0,1\}^n\setminus\{\mathbf{0}\}$ denote a request of a linear combination of information symbols $\sum_{i\in \supp(\mathbf{v})}x_i$, call $\mathbf{v}$ a request-vector. Then, for any multi-set of request vectors 
\begin{equation}\label{eq25}
I=\ms{\mathbf{v}_1^{(a_1)},\ldots,\mathbf{v}_s^{(a_s)}}
\end{equation}
such that $a_i\geq 1$ and $\sum_{i=1}^{s}a_i=t$, there are $t$ mutually disjoint subsets 
$$R_{1,1},\ldots,R_{1,a_1},\ldots, R_{s,1},\ldots,R_{s,a_s}\subseteq[N]$$
such that for every $c\in\cC_F(s,t)$, $1\leq i\leq s$, and $1\leq j\leq a_i$, we have $\sum_{l\in R_{i,j}}c_l=\sum_{l\in \supp(\mathbf{v}_i)}x_l$. We call $R_{i,j}$ a recovering set of $\mathbf{v}_i$.
\end{definition}

When $s=t$, this definition coincides with the definition of functional $t$-batch code in \cite{ZEY20} and when $s=1$, it gives the functional $t$-PIR code. Moreover, the $(s,t)$-batch codes (in binary form) introduced in Definition \ref{GeneralBatch} can be regarded as a specific instance of functional $(s,t)$-batch codes, where each request vector $\mathbf{v}_j$ has weight $1$.

From the definition, a binary functional $(s,t)$-batch code can be represented by an $n\times N$ generator matrix 
$$\mathbf{G}=[\mathbf{g}_1,\ldots,\mathbf{g}_{N}]\in \mathbb{F}_2^{n\times N}$$ 
in which the vector $\mathbf{g}_j$ has $1$'s at positions $i_1,\ldots,i_{l}$ if and only if the $j$-th code symbol $c_j=x_{i_1}+\cdots+x_{i_l}$. 
In this way, a functional $(s,t)$-batch codes can be viewed as an $n\times N$ binary matrix such that for any multi-set of request vectors $I=\ms{\mathbf{v}_1^{(a_1)},\ldots,\mathbf{v}_s^{(a_s)}}$ with $a_i\geq 1$ and $\sum_{i=1}^{s}a_i=t$, there are $t$ mutually disjoint subsets $R_{1,1},\ldots,R_{1,a_1},\ldots, R_{s,1},\ldots,R_{s,a_s}\subseteq [N]$ such that for every $1\leq i\leq s$ and $1\leq j\leq a_i$, the column vectors in $\mathbf{G}$, indexed by $R_{i,j}$, sum up to $\mathbf{v}_i$.

Let $r_F(n;s,t)$ denote the minimum redundancy of a functional $(s,t)$-batch code that encodes $n$ information symbols. Similarly to $(s,t)$-batch codes, for $1\leq s\leq t$, we have
\begin{equation*}\label{eq1f1}
   r_F(n;1,t)\leq r_F(n;s,t)\leq r_F(n;t,t).
\end{equation*}
Moreover, since a functional $(s,t)$-batch code is always an $(s,t)$-batch code, we also have
\begin{equation*}\label{eq1f2}
    r_F(n;s,t)\geq r(n;s,t).
\end{equation*}
If $n$ is clear from the context, we omit $n$ from the notation and write $r_F(s,t)$ instead of $r_F(n;s,t)$.

In previous sections, we focused on $(s,t)$-batch codes for which $1\leq s\leq t\leq n$. 
However, for functional $(s,t)$-batch codes the requests are (non-trivial) linear combinations of information symbols. Thus, the number of distinct request vectors, $\mathbf{v}$, can be $2^n-1$. In this paper, similarly to previous works on functional batch codes (e.g., \cite{ZEY20,YY21}), we assume that $t\leq 2^n-1$ and focus on functional $(s,t)$-batch codes with $1\leq s\leq t\leq 2^n-1$.

Recall that the $[2^{n}-1,n]$ simplex code is a linear code of length $N=2^n-1$ and dimension $n$ whose generator matrix $\mathbf{G}$ comprises of all the non-zero vectors of length $n$. In \cite{ZEY20}, Zhang et al. showed that the $[2^{n}-1,n]$ simplex code is a functional $2^{n-1}$-PIR code, which implies that $r_F(1,2^{n-1})\leq 2^n-n-1$. Moreover, they also proved that $r(1,2^{n-1})=2^n-n-1$. Therefore, by $r_{F}(t,t)\geq r_{F}(1,t)\geq r(1,t)$, we know that $r_{F}(1,2^{n-1})= 2^{n}-n-1$. At the end of \cite{ZEY20}, Zhang et al. conjectured that $r_{F}(2^{n-1},2^{n-1})$ has the same value.
\begin{conjecture}{\cite{ZEY20}}\label{conj1}
$r_{F}(2^{n-1},2^{n-1})=2^n-n-1$ holds for all positive integer $n$. 
\end{conjecture}
This conjecture was verified for the cases $n\leq 5$ in \cite{YKL17} ($n=3,4$) and \cite{ZEY20} ($n=5$) through constructions based on simplex codes. Recently, in \cite{YY21}, the following improvement of the result of \cite{ZEY20} was obtained. 

\begin{theorem}\label{ConstructionFB}\cite{YY21}
For a positive integer $n$, a simplex code of length $2^{n}-1$ and dimension $n$ is also
\begin{itemize}
    \item [a)] a functional $\lfloor\frac{2}{3}\cdot2^{n-1}\rfloor$-batch code; 
    \item [b)] a functional $(\lfloor\frac{5}{6}\cdot2^{n-1}\rfloor-n)$-batch code. 
\end{itemize}
\end{theorem}

Since $r_{F}(1,t)\leq r_{F}(s,t)\leq r_{F}(t,t)$ for all $1\leq s\leq t$, we can rewrite Conjecture \ref{conj1} as follows.
\begin{conjecture}\label{conj2}
$r_{F}(s,2^{n-1})=2^n-n-1$ holds for all positive integer $n$ and $1\leq s\leq 2^{n-1}$. 
\end{conjecture}


In this section, we focus on Conjecture \ref{conj2} for small values of $s$, and obtain the following results.

\begin{theorem}\label{ConstructionGFB}
Let $\mathcal{C}$ be the $[2^{n}-1,n]$ simplex code. Then, for any integer $1\leq s\leq n-\ln{n}$, $\mathcal{C}$ is a functional $(s,2^{n-1}-\lceil\frac{s}{2}\rceil\cdot 2^{s-1})$-batch code. Moreover, $\mathcal{C}$ can answer any multi-set of request vectors $I=\ms{\mathbf{v}_1^{(a_1)},\ldots,\mathbf{v}_s^{(a_s)}}$ such that $\sum_{i=1}^{s}a_i=t$, $\dim(\text{Span}\{\mathbf{v}_1,\ldots,\mathbf{v}_s\})=s'$ and $t\leq2^{n-1}-\lceil\frac{s}{2}\rceil\cdot 2^{s'-1}$.
\end{theorem}

\begin{theorem}\label{ConstructionGFB2}
Let $\mathcal{C}$ be the $[2^{n}-1,n]$ simplex code. Then, $\mathcal{C}$ is a functional $(s,2^{n-1})$-batch code for $s\leq 4$.
\end{theorem}

Since the proofs of Theorem \ref{ConstructionGFB} and Theorem \ref{ConstructionGFB2} require some preliminary results, we defer them to the next section.

\begin{remark}\label{rem5}
Theorem \ref{ConstructionGFB2} confirms Conjecture \ref{conj2} for $s\leq 4$. Theorem \ref{ConstructionGFB} implies that $r_F(s,2^{n-1}(1-o(1)))\leq 2^{n}-n-1$ for $s$ satisfying $s2^{s}=o(2^{n})$, which confirms Conjecture \ref{conj2} asymptotically for $s$ in this parameter regime.
\end{remark}

\subsection{Proofs of Theorem \ref{ConstructionGFB} and Theorem \ref{ConstructionGFB2}}


Let $S=\{\mathbf{v}_1,\ldots,\mathbf{v}_s\}\subseteq\mathbb{F}_2^n$ be a set of $s$ distinct non-zero vectors. Denote by $\langle S \rangle$ the subspace spanned by $\{\mathbf{v}_1,\ldots,\mathbf{v}_s\}$ in $\mathbb{F}_2^{n}$ and let $\dim(S)$ be the dimension of $\langle S\rangle$. The quotient subspace $\mathbb{F}_2^{n}/\langle S\rangle\cong \mathbb{F}_2^{n-\dim(S)}$ induces a partition of $\mathbb{F}_2^{n}$, 
$$\mathbb{F}_2^{n}=\bigcup_{i=1}^{2^{n-\dim(S)}}\mathbf{U}_i=\bigcup_{i=1}^{2^{n-\dim(S)}}\left(\mathbf{u}_i+\langle S\rangle \right),$$
where $\mathbf{u}_i\in\mathbb{F}_2^{n}/\langle S\rangle$ and $\mathbf{U}_i=\mathbf{u}_i+\langle S\rangle$ is an affine subspace of $\mathbb{F}_2^{n}$ of dimension $\dim(S)$. Assume that $\mathbf{u}_1=\mathbf{0}$. Let $\mathbf{U}$ be a subspace of $\mathbb{F}_2^n$ containing $\langle S \rangle$. Then, we define an $s$-partite graph $G_{S,\mathbf{U}}=(V,E)$ as follows:
\begin{itemize}
    \item [a)] The vertex set $V$ consists of $s$ subsets of vertices $V_1,\ldots,V_s$ such that for $1\leq j\leq s$, the set $V_j=\{\mathbf{x}+\langle\mathbf{v}_j\rangle:\mathbf{x}\in \mathbf{U}/\langle\mathbf{v}_j\rangle\}$. In other words, every vertex in $V_j$ corresponds to a 1-dimensional affine subspace $\{\mathbf{x},\mathbf{x+v}_j\}$ for some $\mathbf{x}\in \mathbf{U}/\langle\mathbf{v}_j\rangle$.
    \item [b)] For $1\leq j_1< j_2\leq s$, place an edge between $v_1\in V_{j_1}$ and $v_2\in V_{j_2}$ if $v_1\cap v_2\neq \emptyset$, i.e., if $v_1=\{\mathbf{x}_1,\mathbf{x}_1+\mathbf{v}_{j_1}\}\in V_{j_1}$ and $v_2=\{\mathbf{x}_2,\mathbf{x}_2+\mathbf{v}_{j_2}\}\in V_{j_2}$, then place an edge $(v_1,v_2)\in E$ iff $\{\mathbf{x}_1,\mathbf{x}_1+\mathbf{v}_{j_1}\}\cap \{\mathbf{x}_2,\mathbf{x}_2+\mathbf{v}_{j_2}\}\neq \emptyset$. 
\end{itemize}
Clearly, $V_i$ consists of $2^{\dim(\mathbf{U})-1}$ vertices and each vertex in $V_i$ is a $1$-dim affine subspace translated from $\langle\mathbf{v}_i\rangle$. When $\mathbf{U}=\langle S\rangle$, we abbreviate $G_{S,\mathbf{U}}$ as $G_{S}$.

We obtain the following properties of $G_{S,\mathbf{U}}$. 
\begin{lemma}\label{lem2}
For $1\leq s\leq 2^{n-2}$, let $S=\{\mathbf{v}_1,\ldots,\mathbf{v}_s\}\subseteq \mathbb{F}_2^{n}$ be a set of $s$ distinct non-zero vectors and let $\mathbf{U}$ be a subspace of $\mathbb{F}_2^n$ containing $\langle S \rangle$. Then, $G_{S,\mathbf{U}}$ can be partitioned into $2^{\dim(\mathbf{U})-\dim(S)}$ components $G_{1},\ldots,G_{2^{\dim(\mathbf{U})-\dim(S)}}$. Each $G_i$ is an $s$-partite graph and each part of $G_i$ consists of $2^{\dim(S)-1}$ vertices. Moreover, for every $1\leq i\leq2^{\dim(\mathbf{U})-\dim(S)}$, $G_{i}\cong G_{S}$.
\end{lemma}
\begin{IEEEproof}
We prove the result for the case when $\mathbf{U}= \mathbb{F}_2^{n}$. Since $\mathbf{U}\cong \mathbb{F}_2^{\dim(\mathbf{U})}$, the proof for other cases is similar.

For $1\leq i\leq 2^{n-\dim(S)}$, let $\mathbf{u}_i$ be the $i$-th vector in the quotient subspace $\mathbb{F}_2^{n}/\langle S \rangle$ and define 
$$V_{j,i}=\{\mathbf{x}+\langle\mathbf{v}_j\rangle:\mathbf{x}\in \mathbf{u}_i+ \langle S \rangle/\langle\mathbf{v}_j\rangle\}.$$ 
For each $1\leq j\leq s$, $V_{j,i}= (V_{j}\cap \mathbf{U}_i)$ and $|V_{j,i}|=2^{\dim(S)-1}$. 
Let $G_i$ be the induced subgraph of $G_{S,\mathbf{U}}$ over the vertex set $\bigcup_{j=1}^{s}V_{j,i}$, and let $G_1$ be the subgraph corresponding to $\mathbf{u}_1=\mathbf{0}\in\mathbb{F}_2^{n}/\langle S\rangle$. Then, $G_1=G_{S}$.

Since $\mathbf{U}_1,\ldots,\mathbf{U}_{2^{n-\dim(S)}}$ form a partition of $\mathbb{F}_2^{n}$, we have that for every $1\leq j\leq s$ and $i_1\neq i_2$, $V_{j,i_1}\cap V_{j,i_2}=\emptyset$. For $1\leq i_1\neq i_2\leq 2^{n-\dim(S)}$, define the map $\phi_{i_1,i_2}: \bigcup_{j=1}^{s}V_{j,i_1}\rightarrow \bigcup_{j=1}^{s}V_{j,i_2}$ by 
\begin{align*}
   \phi_{i_1,i_2}\left(\mathbf{x}+\langle\mathbf{v}_j\rangle\right)= \mathbf{x}+\mathbf{u}_{i_2}-\mathbf{u}_{i_1}+\langle\mathbf{v}_j\rangle,~\forall\; 1\leq j\leq s.
\end{align*}
Clearly, $\phi_{i_1,i_2}$ maps vertices from $V_{j,i_1}$ to $V_{j,i_2}$. 
Since $\mathbf{u}_{i_2}-\mathbf{u}_{i_1}\notin \langle S \rangle$ for $i_1\neq i_2$,
$\phi_{i_1,i_2}$ induces a one-to-one correspondence between $V_{j,i_1}$ and $V_{j,i_2}$ for every $1\leq j\leq s$. Moreover, if $\{\mathbf{x}_1+\langle\mathbf{v}_{j_1}\rangle\}\cap \{\mathbf{x}_2+\langle\mathbf{v}_{j_2}\rangle\}\neq \emptyset$, then 
$$(\mathbf{x}_1+\mathbf{u}_{i_2}-\mathbf{u}_{i_1}+\langle\mathbf{v}_{j_1}\rangle)\cap(\mathbf{x}_2+\mathbf{u}_{i_2}-\mathbf{u}_{i_1}+\langle\mathbf{v}_{j_2}\rangle)\neq \emptyset.$$
This implies that if $(\mathbf{x}_1+\langle\mathbf{v}_{j_1}\rangle, \mathbf{x}_2+\langle\mathbf{v}_{j_2}\rangle)$ is an edge in $G_{i_1}$, 
then $(\mathbf{x}_1+\mathbf{u}_{i_2}-\mathbf{u}_{i_1}+\langle\mathbf{v}_{j_1}\rangle,\mathbf{x}_2+\mathbf{u}_{i_2}-\mathbf{u}_{i_1}+\langle\mathbf{v}_{j_2}\rangle)$ is an edge in $G_{i_2}$ and vice-versa. Thus, $\phi_{i_1,i_2}$ actually induces a graph isomorphism between $G_{i_1}$ and $G_{i_2}$.

Now, to complete the proof, we only need to show that there is no edge between $G_{i_1}$ and $G_{i_2}$ for every $1\leq i_1\neq i_2\leq 2^{n-\dim(S)}$. 

Recall that there are no internal edges in $V_j$ for each $1\leq j\leq s$. Assume, towards a contradiction, that there is an edge between $\mathbf{x}_1+\langle\mathbf{v}_{j_1}\rangle\in V_{j_1,i_1}$ and $\mathbf{x}_2+\langle\mathbf{v}_{j_2}\rangle\in V_{j_2,i_2}$ for some $j_1\neq j_2$. Then, we have $(\mathbf{x}_1+\langle\mathbf{v}_{j_1}\rangle)\cap (\mathbf{x}_2+\langle\mathbf{v}_{j_2}\rangle)\neq\emptyset$, which means one of the following cases holds
\begin{equation}\label{eq24}
    \begin{cases}
    \mathbf{x}_1=\mathbf{x}_2;\\
    \mathbf{x}_1=\mathbf{x}_2+\mathbf{v}_{j_2};\\
    \mathbf{x}_1+\mathbf{v}_{j_1}=\mathbf{x}_2;\\
    \mathbf{x}_1+\mathbf{v}_{j_1}=\mathbf{x}_2+\mathbf{v}_{j_2}.
    \end{cases}
\end{equation}
By the definition of $V_{j,i}$, we have 
\begin{equation*}
    \begin{cases}
    \mathbf{x}_1=\mathbf{u}_{i_1}+\mathbf{y}_1\\
    \mathbf{x}_2=\mathbf{u}_{i_2}+\mathbf{y}_2
    \end{cases}
\end{equation*}
for some $\mathbf{y}_1\in  \langle S\rangle/\langle\mathbf{v}_{j_1}\rangle$ and $\mathbf{y}_2\in \langle S\rangle/\langle\mathbf{v}_{j_2}\rangle$. Thus, $\mathbf{x}_1+\mathbf{x}_2=\mathbf{u}_{i_1}+\mathbf{u}_{i_2}+\mathbf{y}_1+\mathbf{y}_2\in \mathbf{u}_{i_1}+\mathbf{u}_{i_2}+\langle S\rangle$, which is not in $\langle S\rangle$. This contradicts (\ref{eq24}), therefore, there is no edge between $G_{i_1}$ and $G_{i_2}$.
\end{IEEEproof}

\begin{lemma}\label{lem3}
For $2\leq s\leq 2^{n-2}$, let $S=\{\mathbf{v}_1,\ldots,\mathbf{v}_s\}\subseteq \mathbb{F}_2^{n}$ be a set of $s$ distinct non-zero vectors with $\dim(S)\geq 2$. Then, for any $s$ non-negative integers $b_1,\ldots, b_s$ such that $\sum_{j=1}^{s}b_j=2^{\dim(S)-2}$, $G_{S}$ has an independent set $A$ of size $2^{\dim(S)-2}$ such that $|A\cap V_j|=b_j$, where $V_j$ is the $j$-th vertex part of $G_S$.
\end{lemma}

\begin{IEEEproof}
We find the desired independent set $A$ through the following greedy approach: Assume that for $1\leq j\leq i-1<s$, we already found a  vertex set $B_j$ from $V_j$ of size $b_j$ such that $\bigcup_{j=1}^{i-1}B_j$ is an independent set in $G_S$. Note that each vertex in $G_S$ has at most $2$ neighbors in one of the other $s-1$ parts. Thus, $\bigcup_{j=1}^{i-1}B_j$ has at most $2\cdot \sum_{j=1}^{i-1}b_j\leq 2^{\dim(S)-1}-b_i$ neighbors in $V_j$. Therefore, we can take $b_i$ vertices from the non-neighbors of $\bigcup_{j=1}^{i-1}B_j$ in $V_j$. Denote these $b_i$ vertices as $B_i$. Then, $\bigcup_{j=1}^{i}B_j$ is an independent set in $G_S$. We can continue this procedure until $i=s$ and obtain an independent set $A=\bigcup_{j=1}^{s}B_j$ such that $|A\cap V_j|=|B_j|=b_j$.
\end{IEEEproof}

Now, armed with Lemma \ref{lem2} and Lemma \ref{lem3}, we proceed with the proof of Theorem \ref{ConstructionGFB}.

\begin{IEEEproof}[Proof of Theorem \ref{ConstructionGFB}]
For integers $1\leq s\leq n-\ln{n}$ and $t=2^{n-1}-\lceil\frac{s}{2}\rceil\cdot 2^{s-1}$, let $I=\ms{\mathbf{v}_1^{(a_1)},\ldots,\mathbf{v}_s^{(a_s)}}$ be an arbitrary multi-set of request vectors such that $a_i\geq 1$ and $\sum_{i=1}^{s}a_i=t$. In the following, to show that the $[2^{n}-1,n]$ simplex code $\mathcal{C}$ is a functional $(s,t)$-batch code, we find $a_j$ recovering sets for each $\mathbf{v}_j$ such that each recovering set has the form $\mathbf{x}+\langle\mathbf{v}_j\rangle=\{\mathbf{x},\mathbf{x}+\mathbf{v}_j\}$ for some $\mathbf{x}\in \mathbb{F}_2^{n}$ and all these $t$ recovering sets are mutually disjoint.

Let $S=\{\mathbf{v}_1,\ldots,\mathbf{v}_s\}$ and $G_{S,\mathbb{F}_2^n}$ be the graph defined by $S$ and $\mathbb{F}_2^n$. 
Define $G_{S,\mathbb{F}_2^n}'$ as the subgraph obtained by deleting from $G_{S,\mathbb{F}_2^n}$ all the edges 
$(\langle\mathbf{v}_{j_1}\rangle, \langle\mathbf{v}_{j_2}\rangle)$ for $1\leq j_1\neq j_2\leq s$. Denote $G'=G_{S,\mathbb{F}_2^n}'$. 
Then, each vertex of $G'$ has the form $\{\mathbf{x},\mathbf{x}+\mathbf{v}_j\}$ for some $\mathbf{x}\in \mathbb{F}_2^n$ and $\mathbf{v}_j\in S$. 
If $G'$ has an independent set $A$ of size $t$ such that $|A\cap V_j|=a_j$ for each $1\leq j\leq s$, then by the definition of the edges in $G'$, we can take the $a_j$ vertices in $A\cap V_j$ as recovering sets for $\mathbf{v}_j$ and all these $t$ recovering sets are mutually disjoint. 
Next, we prove the existence of such an independent set $A$. 

By Lemma \ref{lem2}, $G_{S,\mathbb{F}_2^n}$ can be partitioned into $2^{n-\dim(S)}$ components $G_{1},\ldots, G_{2^{n-\dim(S)}}$. Denote $G_i'=G_i\cap G'$ as the subgraph induced by the vertex sets of $G_i$ in $G'$. For $1\leq i\leq 2^{n-\dim(S)}$, $G_i'$ is an $s$-partite subgraph of $G'$ with $V_{j,i}$ as the $j$-th part of vertex set. Since $G'$ is obtained by removing all the edges of form $(\langle\mathbf{v}_{j_1}\rangle, \langle\mathbf{v}_{j_2}\rangle)$ from $G_{S,\mathbb{F}_2^n}$, for $i_1\neq i_2$, $G_{i_1}'$ and $G_{i_2}'$ remain disconnected in $G'$.

For each $1\leq j\leq s$, write 
$$a_j=\lfloor\frac{a_j}{2^{\dim(S)-1}}\rfloor\cdot 2^{\dim(S)-1}+b_j$$ 
for some $0\leq b_j\leq 2^{\dim(S)-1}-1$ and denote $m=\sum_{j=1}^{s}\lfloor\frac{a_j}{2^{\dim(S)-1}}\rfloor$. 
First, we find an independent set $A_0$ of size $t-\sum_{j=1}^{s}b_j$ such that $|A_0\cap V_j|=\lfloor\frac{a_j}{2^{\dim(S)-1}}\rfloor\cdot 2^{\dim(S)-1}$ through the following process:
\begin{itemize}
    \item First, partition $[m]$ into $s$ parts $B_1,\ldots,B_s$ with $|B_j|=\lfloor\frac{a_j}{2^{\dim(S)-1}}\rfloor$.
    \item Then, take $A_0=\bigcup_{j=1}^{s}\bigcup_{i\in B_j}V_{j,i}$. Since there is no edge between $V_{j_1,i_1}$ and $V_{j_2,i_2}$ for any $i_1\neq i_2$ and $1\leq j_1,j_2\leq s$, $A_0$ is an independent set of $G'$ with size $m2^{\dim(S)-1}=t-\sum_{j=1}^{s}b_j$. 
\end{itemize}
Note that the above process involves the first $m$ components of $G'$. We next pick $\sum_{j=1}^{s}b_j$ vertices from the rest of the components which form an independent set with $A_0$, and complete the construction of $A$. 
Specifically, for each $1\leq j\leq s$ we pick $b_j$ vertices from $\bigcup_{i=m+1}^{2^{n-\dim(S)}}V_{j,i}$. 

By plugging in $t=2^{n-1}-\lceil\frac{s}{2}\rceil\cdot 2^{\dim(S)-1}$ into the equation $m2^{\dim(S)-1}=t-\sum_{j=1}^{s}b_j$, we obtain 
\[\sum_{j=1}^{s}b_j=2^{n-1}-(m+\lceil\frac{s}{2}\rceil)2^{\dim(S)-1}. \]
This implies that $2^{\dim(S)-1}|\sum_{j=1}^{s}b_j$. 
Denoting $M=\frac{\sum_{j=1}^{s}b_j}{2^{\dim(S)-1}}$, we have
\begin{align*}
    2^{n-\dim(S)}-m&= \lceil\frac{s}{2}\rceil+M\\
    &\geq 
    \begin{cases}
    s,~\text{when $2M\geq s$};\\
    2M,~\text{when $2M< s$}.
    \end{cases}
\end{align*} 
We now proceed to the third step:
\begin{itemize}
    \item When $2M\geq s$, we have $m+s\leq 2^{n-\dim(S)}$. Since $b_j\leq 2^{\dim(S)-1}-1$, for each $1\leq j\leq s$ we can take a vertex set $A_j\subseteq V_{j,m+j}$ of size $b_j$. 
    Note that $V_{j,m+j}$ is the $j$-th part of $G_{m+j}'$. Since $G_{i_1}'$ and $G_{i_2}'$ are disconnected for any $i_1\neq i_2$, there are no edges among $A_0,A_1,\ldots,A_s$. 
    Thus, $A=\bigcup_{j=0}^{s}A_j$ is an independent set of $G'$. Specifically, $|A|=|A_0|+|\sum_{j=1}^{s}b_j|=t$ and $|A\cap V_j|=a_j$ for each $1\leq j\leq s$. 
    
    \item When $2M< s$, we have $m+2M\leq 2^{n-\dim(S)}$. Note that $2M\cdot 2^{\dim(S)-2}=\sum_{j=1}^{s}b_j$. 
    Thus, for each $1\leq j\leq s$, there are non-negative integers $b_{j,1},\ldots, b_{j,2M}$ such that $b_j=\sum_{l=1}^{2M}b_{j,l}$, and for every $1\leq l\leq 2M$, $\sum_{j=1}^{s}b_{j,l}= 2^{\dim(S)-2}$. 
    By Lemma \ref{lem3}, for each $1\leq l\leq 2M$, $G_{m+l}'$ has an independent set $A_l$ of size $2^{\dim(S)-2}$ such that $|A_l\cap V_{j,m+l}|=b_{j,l}$ for every $1\leq j\leq s$. 
    Since for any $i_1\neq i_2$, $G_{i_1}'$ and $G_{i_2}'$ are disconnected, there are no edges among $A_0,A_1,\ldots,A_{2M}$. 
    Thus, $A=\bigcup_{l=0}^{2M}A_l$ is an independent set of $G'$ with size $t$ such that $|A\cap V_j|=a_j$ for each $1\leq j\leq s$.
\end{itemize}
This completes the proof.
\end{IEEEproof}

In \cite{YKL17}, the authors confirmed Conjecutre \ref{conj1} for $n=3,4$ using simplex codes of dimension $3$ and $4$. Specifically, for $n=3,4$, they proved that for any set of $s$ ($s\leq 2^{n-1}$) non-zero vectors $S=\{\mathbf{v}_1,\ldots,\mathbf{v}_s\}\subseteq \mathbb{F}_2^{n}$ and positive integers $a_1,\ldots,a_s$ such that $\sum_{i=1}^{s}a_i=2^{n-1}$, the generator matrix $\mathbf{G}$ of the $[2^n-1,n]$ simplex code contains the following $2^{n-1}$ mutually disjoint column sets: 
\begin{equation*}
    \begin{array}{cccc}
       \{\mathbf{v}_1\}, & \{\mathbf{x}_{1,1},\mathbf{x}_{1,1}+\mathbf{v}_1\}, & \cdots & \{\mathbf{x}_{1,a_1-1},\mathbf{x}_{1,a_1-1}+\mathbf{v}_1\},\\
       \{\mathbf{v}_2\}, & \{\mathbf{x}_{2,1},\mathbf{x}_{2,1}+\mathbf{v}_2\}, & \cdots & \{\mathbf{x}_{2,a_2-1},\mathbf{x}_{2,a_2-1}+\mathbf{v}_2\},\\
       \cdots \\
       \{\mathbf{v}_s\}, & \{\mathbf{x}_{s,1},\mathbf{x}_{s,1}+\mathbf{v}_s\}, & \cdots & \{\mathbf{x}_{s,a_s-1},\mathbf{x}_{s,a_s-1}+\mathbf{v}_s\},
    \end{array}
\end{equation*}
where the $\mathbf{x}_{i,j}$'s are distinct, non-zero vectors in $\mathbb{F}_2^{n}$. In view of the graph $G_{S,\mathbb{F}_2^n}'$, we conclude the following.
\begin{lemma}\label{lem4}(\cite{YKL17})
For $1\leq s\leq 2^{n-2}$, let $S=\{\mathbf{v}_1,\ldots,\mathbf{v}_s\}\subseteq \mathbb{F}_2^{n}$ be a set of $s$ distinct non-zero vectors. Then, when $n=3,4$, for any $s$ positive integers $a_1,\ldots, a_s$ such that $\sum_{j=1}^{s}a_j=2^{n-1}$, $G_{S,\mathbb{F}_2^n}'$ has an independent set $A$ of size $2^{n-1}$ such that $|A\cap V_j|=a_j$, where $V_j$ is the $j$-th vertex part of $G_{S,\mathbb{F}_2^n}'$.
\end{lemma}

Next, we prove Theorem \ref{ConstructionGFB2} using Lemma \ref{lem4}.
\begin{IEEEproof}[Proof of Theorem \ref{ConstructionGFB2}]
Since a functional $(s,t)$-batch code is also a functional $(s',t)$-batch code for every $1\leq s'<s$, we only need to show that $\mathcal{C}$ the case when $s=4$.

Let $I=\ms{\mathbf{v}_1^{(a_1)},\ldots,\mathbf{v}_4^{(a_4)}}$ be an arbitrary multi-set of request vectors such that $a_j\geq 1$ and $\sum_{j=1}^{4}a_i=2^{n-1}$. Denote $S=\{\mathbf{v}_1,\ldots,\mathbf{v}_4\}$. Similar to the proof of Theorem \ref{ConstructionGFB}, our aim is to find an independent set $A$ in $G'=G_{S,\mathbb{F}_2^n}'$ of size $2^{n-1}$ such that $|A\cap V_j|=a_j$ for each $1\leq j\leq 4$.

As shown in \cite[Construction 0]{WKCB17}, the $[2^{n}-1,n]$ simplex code can answer any multi-set of request vectors of the form $\ms{\mathbf{e}_1^{(a_1)},\ldots,\mathbf{e}_s^{(a_s)}}$ for any $s\leq n$ and $a_i\geq 1$ satisfying $\sum_{i=1}^{s}a_i=2^{n-1}$. Therefore, through base transformations, $\mathcal{C}$ can answer any multi-set request vectors $I$ such that $\dim(S)=s\leq n$. Moreover, note that any set of $4$ non-zero vectors in $\mathbb{F}_2^{n}$ ($n\geq 3$) has dimension at least $3$. Thus, we only need to consider the case when $\dim(S)=3$. 


For $1\leq i\leq 2^{n-3}$, let $G_i$ be the $i$-th component of $G_{S,\mathbb{F}_2^{n}}$ and denote $G_i'$ as the subgraph induced by the vertex sets of $G_i$ in $G'$. For $1\leq j\leq 4$, denote $V_{j,i}$ as the $j$-th part of vertex set of $G_i$. Then, by Lemma \ref{lem2}, 
$$V_{j,i}=\{\mathbf{x}+\langle\mathbf{v}_j\rangle:\mathbf{x}\in \mathbf{u}_i+ \langle S \rangle/\langle\mathbf{v}_j\rangle\},$$
where $\mathbf{u}_i$ is the $i$-th vector in the quotient subspace $\mathbb{F}_2^{n}/\langle S \rangle$.
Let $\mathbf{U}$ be the $4$-dim subspace of $\mathbb{F}_2^n$ spanned by $\{\mathbf{u}_1=\mathbf{0},\mathbf{u}_{2}\}\cup S$. Then, by Lemma \ref{lem2}, $G_{S,\mathbf{U}}'$ consists of two components $G_1'$ and $G_2'$. 


For each $1\leq j\leq 4$, write 
$$a_j=\lfloor\frac{a_j}{4}\rfloor\cdot 4+b_j$$ 
for some $0\leq b_j\leq 3$. W.l.o.g., assume that $b_1\geq \cdots\geq b_4$. Denote $\tilde{a}_j=\lfloor\frac{a_j}{4}\rfloor$, $m=\sum_{j=1}^{4}\tilde{a}_j$ and $M=\frac{\sum_{j=1}^{4}b_j}{4}$. Since $\sum_{j=1}^{4}a_j=2^{n-1}$, we have $4|\sum_{j=1}^{4}b_j$ and $2^{n-3}=m+M$. Next, we divide the proof into two cases: $M\leq 2$ and $M>2$.

When $M\leq 2$, let $s'$ be the largest index such that $\sum_{j=1}^{s'-1}\tilde{a}_j\leq 2^{n-3}-2$.
For $1\leq j\leq s'-1$, let $B_j=[3+\sum_{l=1}^{j-1}\tilde{a}_l,2+\sum_{l=1}^{j}\tilde{a}_l]$ and let 
$$B_{s'}=[2^{n-3}-\sum_{j=1}^{s'-1}\tilde{a}_j+1,2^{n-3}].$$ 
Then, we find the independent set $A$ through the following steps:
\begin{itemize}
    \item First, take $A_0=\bigcup_{j=1}^{s'}\bigcup_{i\in B_j}V_{j,i}$ and we have $|A_0|=|\bigcup_{j=1}^{s'}B_j|\cdot2^{\dim(S)-1} =2^{n-1}-8$. Note that $G_i'$s are disconnected components of $G'$. Thus, $A_0$ is an independent set of $G'$ such that $|A_0\cap V_j|=a_j-b_j$ for each $1\leq j\leq s'$. Note that the construction of $A_0$ only involves the last $2^{n-3}-2$ components of $G'$.
    \item Second, we find an independent set $A_1$ of size $8=\sum_{j=1}^{s'}b_j+\sum_{j=s'+1}^{4}a_j$ in the first two components $G_1'$ and $G_2'$ of $G'$. Since $\mathbf{U}\cong\mathbb{F}_2^{4}$, Lemma \ref{lem4} implies that for any positive integers $a_1,\ldots,a_4$ such that $\sum_{i=1}^{4}a_i=8$, $G_{S,\mathbf{U}}'=G_1'\cup G_2'$ contains an independent set $A_1$ of size $8$ such that $|A\cap V_j|=a_j$ for each $1\leq j\leq 4$. Thus, $G_1'\cup G_2'$ has an independent set $A_1$ of size $8$ such that $|A_1\cap V_j|=b_j$ for $1\leq j\leq s'$ and $|A_1\cap V_j|=a_j$ for $s'+1\leq j\leq 4$. Therefore, $A=A_0\cup A_1$ is an independent set of $G'$ with size $2^{n-1}$ such that $|A\cap V_j|=a_j$ for each $1\leq j\leq s$.
\end{itemize}

When $M>2$, by $b_j\leq 3$, we have $2<M\leq 3$. Thus, $M=3$ and $b_j=3$ for each $1\leq j\leq 4$. For $1\leq j\leq s$, let 
$$B_j=[4+\sum_{l=1}^{j-1}\tilde{a}_l,3+\sum_{l=1}^{j}\tilde{a}_l].$$ 
Similarly to the previous case, take $A_0=\bigcup_{j=1}^{s}\bigcup_{i\in B_j}V_{j,i}$. Then, $A_0$ is an independent set of $G'$ such that $|A_0\cap V_j|=a_j-3$ for each $1\leq j\leq 4$. Note that $A_0$ is contained in the last $m=2^{n-3}-3$ components of $G'$. Next, we show that $G_1'\cup G_2'\cup G_3'$ contains an independent set $A'$ of size $12$ such that $|A'\cap V_j|=3$ for each $1\leq j\leq 4$. Then, by the disconnectedness among $G_{i}'$s, $A=A_0\cup A'$ is the desired independent set. Since $\dim(S)=3$, w.l.o.g., we can assume that both $\{\mathbf{v}_1,\mathbf{v}_2,\mathbf{v}_3\}$ and $\{\mathbf{v}_2,\mathbf{v}_3,\mathbf{v}_4\}$ are linearly independent. Let 
\begin{align*}
  A_1&=\{\langle\mathbf{v}_1\rangle,\langle\mathbf{v}_2\rangle,\langle\mathbf{v}_3\rangle,\langle\mathbf{v}_4\rangle\},\\
  A_2&=\{\mathbf{u}_2+\langle\mathbf{v}_1\rangle, \mathbf{u}_2+\mathbf{v}_2+\langle\mathbf{v}_1\rangle,\mathbf{u}_2+\mathbf{v}_3+\langle\mathbf{v}_2\rangle,\mathbf{u}_2+\mathbf{v}_1+\mathbf{v}_3+\langle\mathbf{v}_2\rangle\},\\
  A_3&=\{\mathbf{u}_3+\langle\mathbf{v}_3\rangle, \mathbf{u}_3+\mathbf{v}_4+\langle\mathbf{v}_3\rangle,\mathbf{u}_3+\mathbf{v}_2+\langle\mathbf{v}_4\rangle,\mathbf{u}_3+\mathbf{v}_2+\mathbf{v}_3+\langle\mathbf{v}_4\rangle\},
\end{align*}
and $A'=A_1\cup A_2\cup A_3$. Recall that $V_j=\{\mathbf{x}+\langle\mathbf{v}_j\rangle:~\mathbf{x}\in \mathbb{F}_2^{n}/\langle\mathbf{v}_j\rangle\}$. Clearly, $|A'\cap V_j|=3$ for $1\leq j\leq 4$. Moreover, based on the assumption, one can easily check that $A_i$ is an independent set of $G_i'$. Since $G_i'$s are components, $A'$ is an independent set of $G_1'\cup G_2'\cup G_3'$.

This completes the proof.
\end{IEEEproof}

\section{Conclusion and further research}

In this paper, we introduce a generalization of PIR codes and batch codes - the $(s,t)$-batch codes. We study the trade-off between the redundancy and the required $(s,t)$-batch repair property of this new class of codes. First, we proved a lower bound on the redundancy for general $s$ and $t$ which coincides with the result of \cite{LW21batch,AG21} for the case $s=t$ and slightly improves the result of \cite{RVW22} for the case $s=1$. Then, we provide a recursive construction of $(s,t)$-batch codes. Moreover, based on a random construction, we prove the existence of $(s,t)$-batch code with redundancy $O(t^{3/2}\sqrt{n})$ for $s=O(\frac{t}{\ln{n}})$ and $t=o(n^{1/3})$. Finally, we considered functional $(s,t)$-batch codes and show that simplex codes are asymptotically optimal with respect to the redundancy for relatively small $s$.

As new variants of PIR codes and batch codes, we believe that the $(s,t)$-batch codes (functional $(s,t)$-batch codes) introduced in this paper offer more flexible parameters for practical scenarios. Moreover, from the theoretical point of view, the $(s,t)$-batch code and functional $(s,t)$-batch code can be good starting points for finding the differences between PIR codes and batch codes, as well as between functional PIR codes and functional batch codes. 

In the following, we list some questions for further research on this topic.

\begin{enumerate}    
    \item In the proof of Theorem \ref{thm1}, the field size is required to be polynomial in $n$. Can this constraint be removed? Or can the lower bound be improved?
    \item For general $s<t$, comparing with the result of $t$-batch codes, the constructions in Section \ref{sec_IVb} only provides a $\ln{n}$ factor improvement for the case $s=O(\frac{t}{\ln{n}})$ and $t=o(n^{1/3})$ and this construction is probabilistic. It is interesting to provide an explicit construction of $(s,t)$-batch codes for $s=o(t)$ with essentially smaller redundancy than $t$-batch codes.
    \item As in \cite{ZEY20,YY21}, we also believe that Conjecture \ref{conj2} indeed holds true by using simplex codes. More specifically, we believe that the $s$-partite graph $G_{S,\mathbb{F}_2^n}'$ defined in Section V.B contains an independent set with $a_j$ vertices in each part for any $(a_1,\ldots,a_s)\in \mathbb{F}_2^{s}$ satisfying $\sum_{j=1}^{s}a_j=2^{n-1}$. 
\end{enumerate}

\section*{Acknowledgement}

We thank Prof. Itzhak Tamo for his many illuminating discussions and especially, for sharing the idea of the proof of Theorem \ref{thm1} for the case when $u=1$. 
We thank Prof. Yiwei Zhang for introducing works \cite{WKCB17} and \cite{ZEY20} to us.

\appendices

\section{Counting subspaces}

For positive integers $k>s\geq 1$, let $U$ be a $k$-dim subspace of $\mathbb{F}_q^{n}$ and $U_1$ be a $(k-s)$-dim subspace of $U$. It's a natural question to ask the number of $l$-dim subspaces in $\mathbb{F}_q^{n}$ whose intersection with $U$ is contained in $U_1$. As shown in \cite[Lemma 9.3.2]{GM15}, the number of $l$-dim subspaces whose intersection with $U$ has dimension $j$ is given by $q^{(k-j)(l-j)}\qbinom{n-k}{l-j}_q\qbinom{k}{j}_q$, where $\qbinom{n}{k}_q = \frac{(q^n-1)\dots(q^{n-k+1}-1)}{(q^k-1)\dots (q-1)}$. To solve the above question, we generalize this result as follows.

\begin{lemma}\label{lem71}
The number of $l$-dim subspaces whose intersection with $U$ is contained in $U_1$ and has dimension $j$ is given by $$q^{(k-j)(l-j)}\qbinom{n-k}{l-j}_q\qbinom{k-s}{j}_q.$$
\end{lemma}

For the proof of Lemma \ref{lem71}, we need the following result.

\begin{lemma}\label{lem72}\cite[Lemma 9.3.1]{GM15}
The number of $l$-dim subspaces of a $(k+l)$-dim space,  which intersect a fixed $k$-dim subspace only at $\{0\}$ is $q^{kl}$.
\end{lemma}

\begin{IEEEproof}[Proof of Lemma \ref{lem71}]
Let $\mathcal{M}$ denote the set of $l$-dim subspaces whose intersection with $U$ is $\{\mathbf{0}\}$. Deem two elements $V_1$ and $V_2$ of $\mathcal{M}$ to be equivalent if 
$$\text{Span}_{\mathbb{F}_q}\{U,V_1\}=\text{Span}_{\mathbb{F}_q}\{U,V_2\}.$$
Note that $\text{Span}_{\mathbb{F}_q}\{U,V_1\}$ (or $\text{Span}_{\mathbb{F}_q}\{U,V_2\}$) is a $(k+l)$-dim subspace of $\mathbb{F}_q^{n}$. Therefore, the subspaces of  dimension $k+l$ that contain $U$ partition the elements of $\mathcal{M}$. By Lemma \ref{lem72}, the number of subspaces in a class of this partition is $q^{kl}$, and the number of classes is the number of different $(k+l)$-dim subspaces containing $U$ in $\mathbb{F}_q^n$. Therefore, 
$$|\mathcal{M}|=q^{kl}\qbinom{n-k}{l}_q.$$

It follows that the number of $l$-dim subspaces $V$ such that $U\cap V$ is a fixed $j$-dim subspace of $U$ equals to
$q^{(k-j)(l-j)}\qbinom{n-k}{l-j}_q$. Note that there are $\qbinom{k-s}{j}_q$ such $j$-dim subspaces in $U_1$. Therefore, the number of $l$-dim subspaces whose intersection with $U$ is contained in $U_1$ and has dimension $j$ is
$$q^{(k-j)(l-j)}\qbinom{n-k}{l-j}_q\qbinom{k-s}{j}_q.$$
\end{IEEEproof}

\section{Some inequalities of $q$-binomial coefficients}

In this part, we prove several inequalities about $q$-binomial coefficients.

\begin{lemma}\label{lem73}
Let $q$ be a prime power. For positive integers $n\geq k\geq 1$, we have the following inequalities:
\begin{itemize}
    \item [1.] $\qbinom{n}{k}_q\geq q^{k(n-k)}$ and an equality holds if and only if $n=k$.
    \item [2.] $\qbinom{n}{k}_q<q^{k(n-k)}(1+\frac{2k}{q-1})$.
\end{itemize}
\end{lemma}
\begin{IEEEproof}
The first inequality follows from $\qbinom{n}{k}_q = \frac{(q^n-1)\cdots(q^{n-k+1}-1)}{(q^k-1)\cdots (q-1)}$ and $\frac{q^{n-i}-1}{q^{k-i}-1}\geq q^{n-k}$ for all $0\leq i< k$, where the equality holds if and only if $n=k$.

We use induction on $k$ to prove the second inequality. When $k=1$, we have 
$$\qbinom{n}{1}_q=\frac{q^{n}-1}{q-1}<\frac{q^{n}+q^{n-1}}{q-1}=q^{n-1}(1+\frac{2}{q-1}).$$
Now assume that $\qbinom{n}{l}_q<q^{l(n-l)}(1+\frac{2l}{q-1})$ holds for all $n$ and $1\leq l\leq k-1$. Since $\qbinom{n}{k}_q=\frac{q^{n}-1}{q^k-1}\cdot \qbinom{n-1}{k-1}$, by induction hypothesis, we have
\begin{align*}
    \qbinom{n}{k}_q &< \frac{q^{n}-1}{q^k-1}\cdot q^{(k-1)(n-k)}\cdot(1+\frac{2k-2}{q-1})\\
    &= q^{k(n-k)}\cdot(1+\frac{2k-2}{q-1})\cdot (1+\frac{1-q^{k-n}}{q^k-1})\\
    &< q^{k(n-k)}\cdot(1+\frac{2k}{q-1}),
\end{align*}
where the last inequality follows from $2k-2<q^k-1$.
\end{IEEEproof}

\section{Proof of Lemma \ref{lem:chern}}

\begin{IEEEproof}   
    The proof of the second inequality 
    \[\Pr(X\leq (1-\delta)\mu)\leq e^{-\mu\delta^2/2},\qquad \forall~ 0<\delta<1\] 
    can be found in \cite[Thm. 4.5]{Prob2017}.
    For the first inequality, we start with the bound as appears in \cite[Thm. 4.4]{Prob2017}. 
    With notation as in the lemma, 
    \[\Pr(X\geq (1+\delta)\mu)\leq \parenv{\frac{e^{\delta}}{(1+\delta)^{1+\delta}}}^{\mu},\qquad \forall~ \delta>0.\] 
    Notice that 
    \begin{align}
    \label{eq:me1}
    \log (1+\delta)&\stackrel{(a)}{\geq} \delta-\frac{1}{2}\delta^2=\frac{\delta(2-\delta)}{2} \nonumber \\ 
    &= \frac{\delta(4-\delta^2)}{2(2+\delta)}\stackrel{(b)}{\geq}\frac{2\delta}{2+\delta},
    \end{align}
    where $(a)$ follows from taking second-order Taylor approximation and noticing that the reminder term has a positive sign, and $(b)$ follows since $\delta>0$. 
    Moreover, we have  
    \begin{align*}
        \log (1+\delta)^{(1+\delta)\mu}&=(1+\delta)\mu\log (1+\delta)\\ 
        &\stackrel{\text{Eq. }(\ref{eq:me1})}{\geq} \frac{(1+\delta)\mu2\delta}{2+\delta}.
    \end{align*}
    Thus, 
    \[(1+\delta)^{(1+\delta)\mu}\geq e^{\frac{(1+\delta)\mu2\delta}{2+\delta}},\] 
    which, in turn, implies 
    \[\parenv{\frac{e^{\delta}}{(1+\delta)^{1+\delta}}}^{\mu}\leq \frac{e^{\delta \mu}}{e^{\frac{(1+\delta)\mu2\delta}{2+\delta}}}.\]
    Simplifying the right-hand-side, we get  
    \begin{align*}
        \frac{e^{\delta \mu}}{e^{\frac{(1+\delta)\mu2\delta}{2+\delta}}}&= e^{\delta \mu-\frac{(1+\delta)\mu2\delta}{2+\delta}}\\ 
        &= e^{\frac{-\delta^2\mu}{2+\delta}}.
    \end{align*}
    Plugging this into the Chernoff bound finishes the proof.
\end{IEEEproof}

\bibliographystyle{IEEEtran}
\bibliography{biblio}

\end{document}